\documentclass[aps,prd,twocolumn,superscriptaddress,amssymb,eqsecnum,showpacs,showkeyes,secnumarabic,graphics,floatfix,nofootinbib,tightenlines,longbibliography]{revtex4-1}

\usepackage{graphicx}
\usepackage{epsfig}
\usepackage{bm}
\usepackage{cancel}
\usepackage{mathtools}
\usepackage{dcolumn}
\usepackage{amsmath}
\usepackage{hhline}
\usepackage{longtable}
\usepackage[utf8]{inputenc}
\usepackage[breaklinks=true,colorlinks=true,
linkcolor=blue,urlcolor=blue,citecolor=cyan,
bookmarks=true,bookmarksopenlevel=2]{hyperref}

\def \beq  {\begin{equation}}
\def \eeq  {\end{equation}}
\def \ber  {\begin{eqnarray}}
\def \eer  {\end{eqnarray}}

\begin{document}
\newcommand{\newc}{\newcommand}

\newc{\be}{\begin{equation}}
\newc{\ee}{\end{equation}}
\newc{\ba}{\begin{eqnarray}}
\newc{\ea}{\end{eqnarray}}
\newc{\bea}{\begin{eqnarray*}}
\newc{\eea}{\end{eqnarray*}}
\newc{\D}{\partial}
\newc{\ie}{{\it i.e.} }
\newc{\eg}{{\it e.g.} }
\newc{\etc}{{\it etc.} }
\newc{\etal}{{\it et al.}}
\newc{\lcdm}{$\Lambda$CDM}
\newcommand{\nn}{\nonumber}
\newc{\ra}{\Rightarrow}

\date{\today}
\title{Constraints on Spatially Oscillating Sub-mm Forces from the Stanford Levitated Microsphere Experiment Data}

\author{I. Antoniou}\email{ianton@uoi.gr}
\affiliation{Department of Physics, University of Ioannina,
GR-45110, Ioannina, Greece}
\author{L. Perivolaropoulos}\email{leandros@uoi.gr}
\affiliation{Department of Physics, University of Patras, 26500 Patras, Greece 
(on leave from the Department of Physics, University of Ioannina, 45110 Ioannina, Greece)}

\begin{abstract}
A recent analysis by one of the authors\cite{Perivolaropoulos:2016ucs} has indicated the presence of a $2\sigma$ signal of spatially oscillating new force residuals in the torsion balance data of the Washington experiment.  We extend that study and analyse the data of the Stanford Optically Levitated Microsphere Experiment (SOLME)  \cite{Rider:2016xaq} (kindly provided by the authors of \cite{Rider:2016xaq}) searching for sub-mm spatially oscillating new force signals. We find a statistically significant oscillating signal for a force residual of the form $F(z)=\alpha \; cos(\frac{2\pi}{\lambda}\; z +c)$ where $z$ is the distance between the macroscopic interacting masses (levitated microsphere and cantilever).  The best fit parameter values are $\alpha=(1.1 \pm 0.4)\times 10^{-17}N$, $\lambda=(35.2\pm 0.6)\mu m$. Monte Carlo simulation of the SOLME data under the assumption of zero force residuals has indicated that the statistical significance of this signal is at about $2\sigma$ level. The improvement of the $\chi^2$ fit compared to the null hypothesis (zero residual force) corresponds to $\Delta \chi^2 = 13.1$. Private communication with the authors of Ref. \cite{Rider:2016xaq} has indicated that this previously unnoticed signal is indeed in the data but is most probably induced by a systematic effect caused by diffraction of non-Gaussian tails of the laser beam. Thus the amplitude of this detected signal can only be useful as an upper bound to the amplitude of new spatially oscillating forces on sub-mm scales. In the context of gravitational origin of the signal emerging from a fundamental modification of the Newtonian potential of the form $V_{eff}(r)=-G\frac{M}{r}(1+\alpha_O \cos(\frac{2\pi}{\lambda}\; r+\theta))\equiv V_{N}(r)+V_{osc}(r)$, we evaluate the source integral of the oscillating macroscopically induced force. If the origin of the SOLME oscillating signal is systematic, the parameter $\alpha_O$ is bounded as $\alpha_O < 10^7$ for $\lambda \simeq 35 \mu m$. Thus, the SOLME data can not provide useful constraints on the modified gravity parameter $\alpha_O$. However, the constraints on the general phenomenological parameter $\alpha$ ($\alpha < 0.3 \times 10^{-17}N$ at $2\sigma$) can be useful in constraining other fifth force models related to dark energy (chameleon oscillating potentials etc). 
\end{abstract}
%
%
\maketitle

\section{Introduction}
\label{sec:Introduction}

The physical scale associated with the accelerating expansion of the universe is the dark energy scale which is obtained from the  dark energy density  $\rho_{de} \simeq 10^{-29}g/cm^3 \simeq (2.4 meV)^4$. This scale corresponds to an energy scale of $2.4meV$ and a length scale of about $\lambda_{de}\simeq \sqrt[4]{\frac{\hbar c}{\rho_{de}}}=0.085mm$. It is therefore plausible that the physical cause of the cosmological expansion on cosmological scales may also produce experimental signatures in the form of new forces that manifest themselves on sub-mm scales. Chameleon scalar field screened interactions \cite{Brax:2004qh,Khoury:2003aq,Brax:2008hh,Gannouji:2010fc,Khoury:2013yya,Burrage:2016bwy,Khoury:2003rn,Antoniou:2015sga}, modified gravity Yukawa forces \cite{Koyama:2015vza,Rahvar:2014yta} and vacuum energy Casimir forces \cite{Bordag:2001qi,Klimchitskaya:2009cw,Milton:2004ya} are some examples of new sub-mm forces that could also be connected with the observed cosmological accelerating expansion. 

A wide range of experiments have been performed searching for signatures of new forces on sub-mm scales.  They include  torsion balance experiments\cite{Kapner:2006si,Fischbach:1999bc,Adelberger:1992ph,Adelberger:2003zx,Chiaverini:2002cb,Long:2002wn,Hoskins:1985tn,Spero:1980zz,Moody:1993ir,Kuroda:1985cyy,Chan:1982zz,Adelberger:2009zz,Smullin:2005iv,Geraci:2008hb,Ninomiya:2017gwv}, Casimir force experiments \cite{Lamoreaux:1996wh,Barvinsky:2011hd,Klimchitskaya:2017cnn}, levitating microsphere experiments \cite{Rider:2016xaq,Li:2011ju,Geraci:2010ft,Moore:2014yba,Wang:2016ngs}, atomic interferometry \cite{Brax:2016wjk}  etc. These experiments as well as astrophysical observations on larger scales \cite{Hees:2017aal} fit particular parametrizations to datasets that usually involve force or torque residuals as function of separation between interacting bodies.

Parametrizations that are commonly used to model the spatial dependence of new forces on sub-mm scales are monotonic and include Yukawa  and power law parametrizations \cite{Adelberger:2006dh}. Yukawa parametrizations generalize the gravitational potential generated by a mass $M$ to the  form
\be V_{eff}=-G\frac{M}{r}(1+\alpha_Y e^{-r/\lambda})\label{vyuk}\ee  
where $\alpha_Y$, $\lambda$ are parameters to be constrained by the data.  Power law parametrizations generalize the gravitational potential generated by a mass $M$ to the  form

\be 
V_{eff}= -G \frac{M}{r}(1+\beta^k(\frac{\lambda}{r})^{k-1})
\label{poweranz}
\ee
where $\beta$, $k$ are parameters.
These parametrizations are motivated by viable extensions of General Relativity (Brans-Dicke and scalar-tensor theories \cite{Perivolaropoulos:2009ak,Hohmann:2013rba,Jarv:2014hma} brane world modes \cite{Nojiri:2002wn,Donini:2016kgu,Benichou:2011dx,Bronnikov:2006jy,Guo:2014bxa,Kaminski:2009dh}, $f(R)$ theories \cite{Berry:2011pb,Capozziello:2009vr,Schellstede:2016ldu}, compactified extra dimension models\cite{ArkaniHamed:1998rs,ArkaniHamed:1998nn,Antoniadis:1998ig,Perivolaropoulos:2002pn,Floratos:1999bv,Kehagias:1999my,Antoniadis:2015jzk}.  Alternative more complicated parametrizations which may not appear in closed analytic form are obtained in the context of non-relativistic, steady-state chameleon fields, that couple directly to matter density and can mediate screened new forces between macroscopic objects \cite{Mota:2006ed,Burrage:2014oza,Elder:2016yxm,Rider:2016xaq} which may even be significantly larger than gravity \cite{Mota:2006ed}. 

Recent studies \cite{Edholm:2016hbt,Conroy:2014eja,Perivolaropoulos:2016ucs,Conroy:2017nkc} have pointed out that a new class of parametrizations describing spatially oscillating new forces on sub-mm scales is well motivated theoretically and viable experimentally. Such oscillating parametrizations may describe deviations of the gravitational force from a Newtonian force in a wide range of modified gravity theories\cite{Perivolaropoulos:2016ucs}, in theories involving small scale granularity of dark energy\cite{Burikham:2017bkn,Lake:2017ync} and most importantly in non-local (infinite derivative) gravity theories \cite{Edholm:2016hbt,Modesto:2017sdr,Amendola:2017qge,Perivolaropoulos:2016ucs,Conroy:2014eja,Rahvar:2014yta,Conroy:2017nkc,Conroy:2017uds,Koshelev:2017bxd,Calcagni:2017sov,Mazumdar:2017kxr}. These theories can be free from singularities \cite{Conroy:2017uds,Koshelev:2017bxd,Buoninfante:2016iuf,Giacchini:2016xns} (such as black holes) and instabilities\cite{Mazumdar:2017kxr,Calcagni:2017sov,Accioly:2016etf}, they can emerge from quantum effects\cite{Amendola:2017qge} (such as light particle loops) and they do not need the existence of the cosmological constant $\Lambda$ to interpret the cosmological observations\cite{Dirian:2014bma}. They constitute a viable physical mechanism for the observed accelerating expansion of the universe \cite{Dirian:2016puz,Maggiore:2014sia,Vardanyan:2017kal,Park:2016jym} while they predict specific signatures in the gravitationally light bending angle \cite{Feng:2017vqd} testable by the Chandra X-ray Observatory\footnote{http://chandra.harvard.edu/}. 

Oscillating force residuals are experimentally viable and mildly favored \cite{Perivolaropoulos:2016ucs} according to current torsion balance experiments searching for new forces on sub-mm scales \cite{Kapner:2006si}. These parametrizations also emerge as analytic continuations of the Yukawa parametrization (\ref{vyuk}) and generalize the Newtonian gravitational potential as
\be V_{eff}=-G\frac{M}{r}(1+\alpha_O \cos(\frac{2\pi}{\lambda} r+\theta))\equiv V_{N}(r)+V_{osc}(r)
\label{vosc}
\ee 
where $\alpha_O$, $\lambda$, $\theta$ are free parameters and the spatial wavelength $\lambda$ is assumed to be of sub-mm scale for consistency with current experimental constraints. This type of parametrization leads to oscillating new forces of sub-mm wavelength of the form 
\begin{widetext}
\be 
{\vec F}=-{\hat r} \frac{GMm}{r^2}\left[1+\alpha_O \cos(\frac{2\pi}{\lambda} r+\theta)+\alpha_O \frac{2\pi}{\lambda} r \sin(\frac{2\pi}{\lambda} r +\theta)\right]
\label{oscforce}
\ee
\end{widetext}
In the case of interacting macroscopic bodies the gravitational potential energy (and therefore the gravitational force) can be obtained by integration of the oscillating potential energy correction term (source integral obtained from the potential $V_{osc}$ eq. (\ref{vosc})) over the volumes of the interacting bodies. Assuming macroscopic interacting masses $M$ and $m$ with a common density $\rho$, the corresponding potential energy source integral may be written as
\begin{widetext}
\be V_{osc}(r)=-G\alpha_O \int_{V_{m}}d^3r_{m}\rho(\vec{r}_{m}) \int_{V_{M}}d^3r_{M}\rho(\vec{r}_{M})\frac{\cos(\frac{2\pi}{\lambda}|\vec{r}_{m}-\vec{r}_{M}|+\theta)}{|\vec{r}_{m}-\vec{r}_{M}|}\label{potsourceint}\ee
\end{widetext}

As demonstrated in section \ref{sec:sourceint} the effective force obtained from the potential source integral (\ref{potsourceint}) macroscopic cylinder of mass $M$ interacting with a small mass $m$ located at a distance $z$ from one of its bases along its symmetry axis is well approximated for intermediate to large $z$ as  
\be 
{\vec F}(z)=A \cos(\frac{2\pi}{\lambda}z+c){\hat z}
\label{oscforce1}
\ee
where $c$ is a parameter. Oscillating sub-mm force residuals like (\ref{oscforce1}) were shown in Ref. \cite{Perivolaropoulos:2016ucs} to be consistent with current torsion balance experiments \cite{Kapner:2006si} and in fact to provide a somewhat better fit than the null hypothesis of zero force residuals. 

In the present study we extend the analysis of Ref \cite{Perivolaropoulos:2016ucs} by fitting the spatially oscillating force residual  (\ref{oscforce1}) to the dataset of Stanford Optically Levitated Microsphere Experiment (SOLME)  \cite{Rider:2016xaq} involving force measurements on  optically levitated microspheres  as a function of its distance $z$ from a gold coated silicon cantilever. The residual force obtained after the subtraction of a best fit electrostatic background from the total measured force in units of $fN$ for $z\in [25,235]\mu m$ is fit to the oscillating force residual parametrization of eq. (\ref{oscforce1}) and the quality of fit is compared to the null hypothesis of zero force residual. The analytic expression of the source integral (\ref{potsourceint}) is also investigated and its quality of fit to the SOLME data is compared to the corresponding quality of fit of the simpler approximate form (\ref{oscforce1}) and other monotonic parametrizations.

This work is organized as follows: in section II we  
describe  the Stanford Optically Levitated Micropshere Experiment (SOLME) \cite{Rider:2016xaq} and the dataset used in our analysis. We also present the analysis of the dataset and compare the likelihood of the oscillating residual force parametrization (\ref{oscforce1}) with the likelihood of the null hypothesis (absence of any residual force). In section III we derive analytical expression for the source integrals leading to the macroscopic residual forces corresponding to the potential between a cylinder and a small sphere. We compare the quality of fit (likelihood) of macroscopic Yukawa, oscillating  and power law residuals. Finally, in section IV we conclude, summarize our results and discuss possible prospects of the present work.

\section{Constraints on a Phenomenological Oscillating Parametrization}
\label{sec:Section1}
The SOLME \cite{Rider:2016xaq} uses optically levitated dielectric microspheres supported by the radiation pressure from a single upward pointing laser beam. The laser traps the microsphere in a high vacuum thus counterbalancing Earth gravity. Any additional force is assumed to be due to a gold-coated silicon cantilever, located in the same height with the microsphera. In order to minimize electrostatic background forces the trap and cantilever are shielded in a cubic container consisting of  six gold-plated
electrodes which are set to approximately equal potential as the cantilever. Despite of this shielding, the main background force remains of electrostatic origin. It emerges due to the interaction of the small but non-zero permanent electric dipole moment of the micro-spheres which couples to the electric field due to the small but non-zero potential difference fluctuations ($<30mV$) between the cantilever and shielding electrodes. Thus the best fit electrostatic background may be used to obtain the residual force data  as the difference between the measured total force and the best fit electrostatic background force at a given microsphere-cantilever distance $z$. Thus for the magnitude of the residual force $dF$ we have 

\be dF\equiv F_{measured}-F_{background}\label{df}\ee

\begin{figure}[b]
\centering
\vspace{0cm}\rotatebox{0}{\vspace{0cm}\hspace{0cm}\resizebox{0.49\textwidth}{!}{\includegraphics{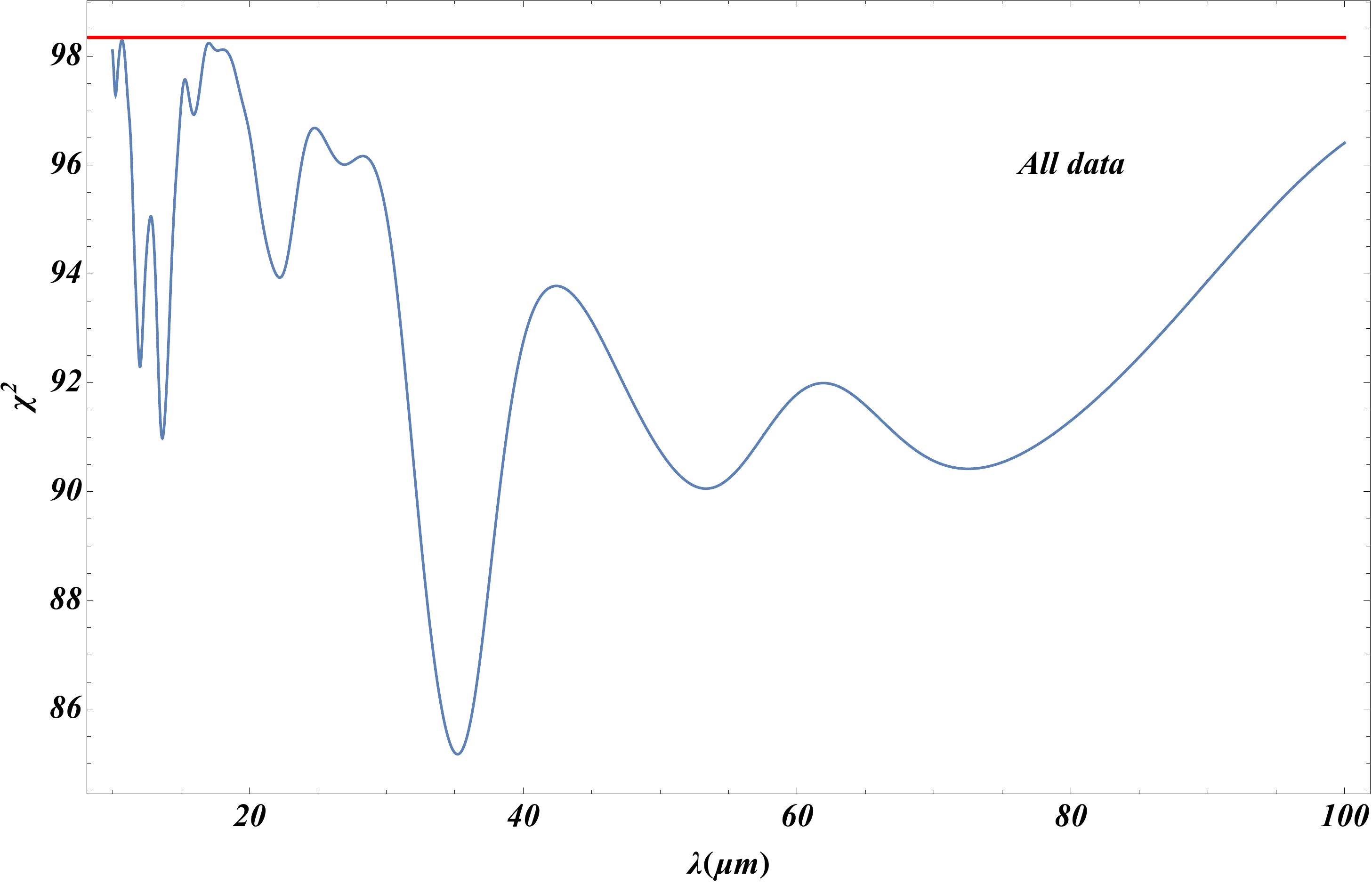}}}
\caption{\small The value of the minimized $\chi^2$ as a function of the wavelength $\lambda$ for the full dataset ($96$ points). The red straight line corresponds zero residual force $dF=0$. The depth of the minimum is $\delta\chi^2=13.1$.}
\label{chi2all}
\end{figure}
 
 The dataset analyzed in the present study corresponds to the data shown in Fig. $3$ of Ref. \cite{Rider:2016xaq}. The data and the best fit electrostatic background were kindly provided by the members of the SOLME  \cite{Rider:2016xaq} after our request.  This dataset was obtained using three silica microspheres with the same radius $r=2.5 \mu m$ and mass $m=0.13ng$ but different polarizabilities. Each microsphera was trapped in a high vacuum with pressure $P<10^{-6}mbar$ and its position was measured by a position-sensitive photodiode using a laser beam. 

\begin{figure}[b]
\centering
\vspace{0cm}\rotatebox{0}{\vspace{0cm}\hspace{0cm}\resizebox{0.49\textwidth}{!}{\includegraphics{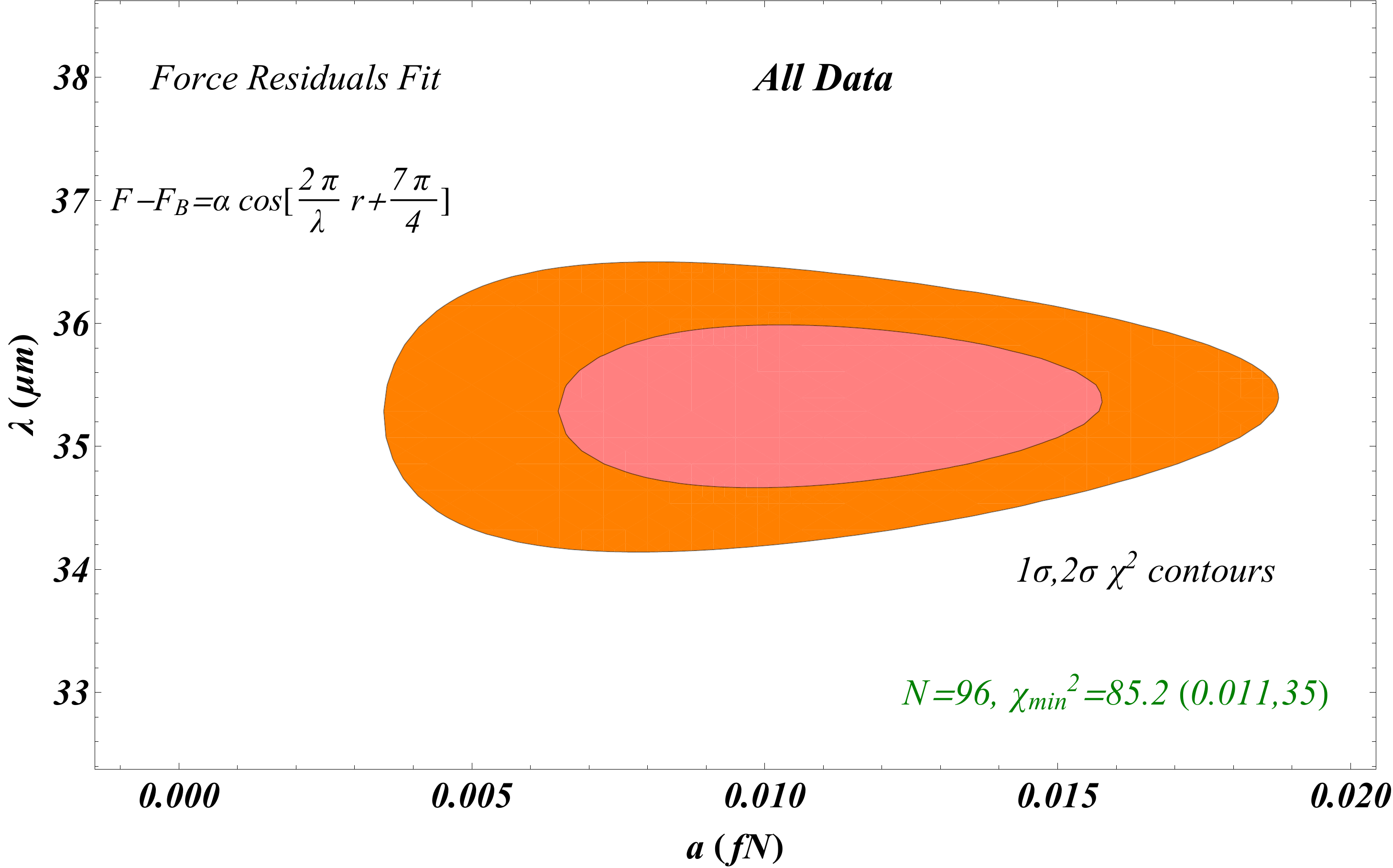}}}
\caption{\small The $1\sigma$ and $2\sigma$ contours in the parameter space $(\alpha,\lambda)$  for the oscillating parametrization with $c=7 \pi /4$. For the combined dataset ($96$ datapoints) there is a well defined high quality fit at $(\alpha,\lambda)=(0.011,35.2 \mu m)$ corresponding to a wavelength $\lambda=35.2\mu m$. This best fit is about $3\sigma$ away from the zero force residual value $\alpha=0$.}
\label{contourall}
\end{figure}

The small unshielded electrostatic background forces are monotonic with the distance $z$ between cantilever and microsphera and have been modelled and fit by the members of the SOLME as functions of the distance $z$ between the cantilever and the microsphere. We have found that this background is very well fit by a parametrization of the form $F_B=a+b/r^{3/2}$ where $a$, $b$ are appropriate parameters that depend on the polarizabity of the interacting micropshere. Any new type of force would manifest itself as a statistically significant nonzero residual force beyond the modelled electrostatic background. 

\begin{figure}[b]
\centering
\vspace{0cm}\rotatebox{0}{\vspace{0cm}\hspace{0cm}\resizebox{0.49\textwidth}{!}{\includegraphics{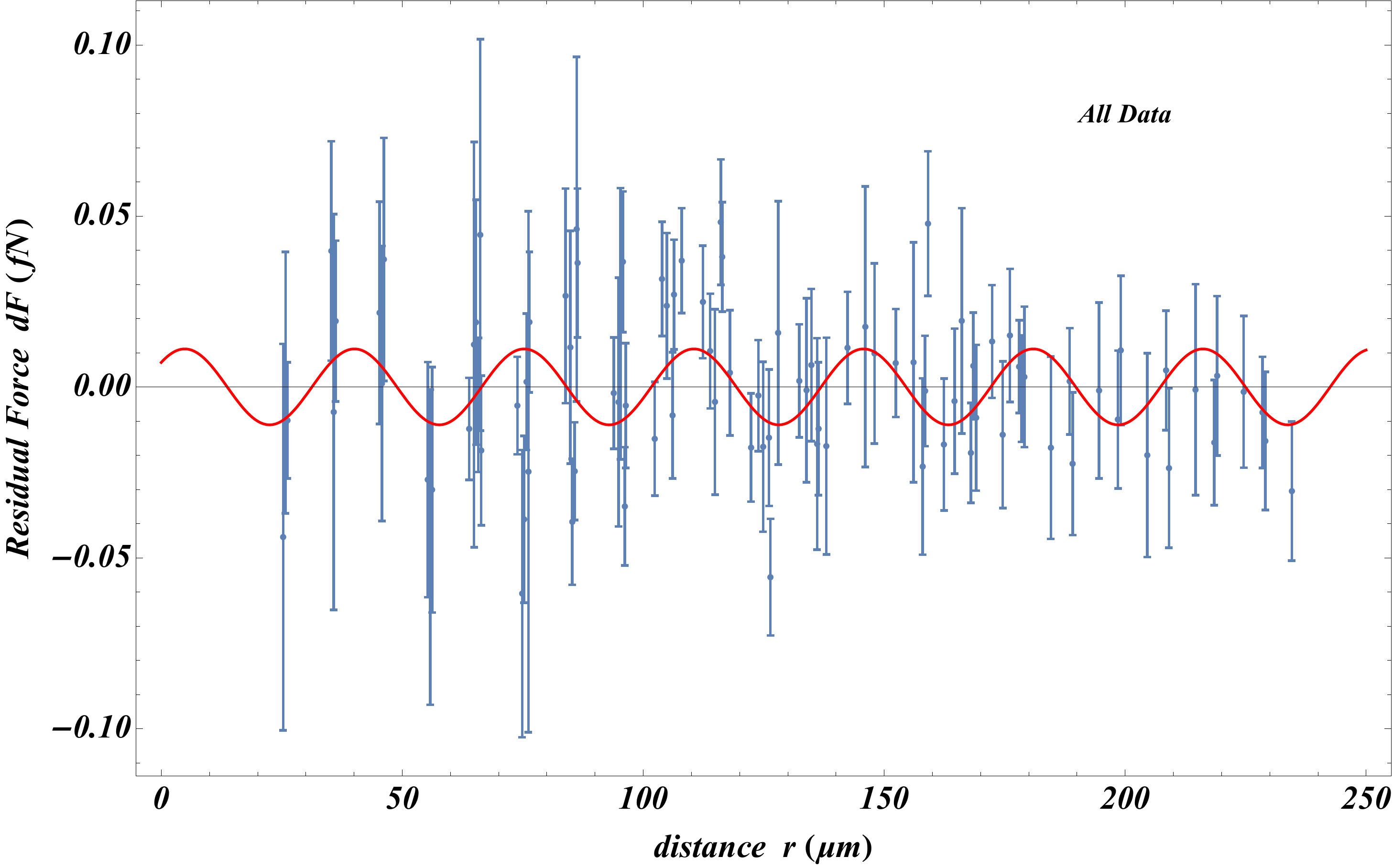}}}
\caption{\small The residual force SOLME data with error bars along with the best fit oscillating parametrization (thin red line) for the full dataset. The best fit harmonic parametrization has spatial wavelength $\lambda=35.2\mu m$.}
\label{bfit}
\end{figure}

For each one of the three silica microsphere the residual force of eq. (\ref{df}) was obtained for $32$ distances $z$ between cantilever and microsphere in the distance range $z$ from $25\mu m$ up to $\sim235\mu m$. The total of $96$ values of these residual forces along with the corresponding distances $z$ and their $1\sigma$ error is shown in Table I in the Appendix (32 values for each one of the three microsphere).

We fit the residual forces of the SOLME data derived from equation (\ref{df}) using the oscillating parametrization of the form 

\be dF(\alpha,\lambda,c,z)=\alpha \;cos(\frac{2\pi}{\lambda}\; z+c)\label{model}\ee

\noindent where $\alpha$, $\lambda$ and $c$ are parameters to be fit.
We have used the parametrization (\ref{model}) to minimize $\chi^2(\alpha,\lambda,c)$ defined as \be 
\chi^2(\alpha,\lambda,c)=\sum_{j=1}^{N}\frac{\left(dF(j)-dF(\alpha,\lambda,c,z_j)\right)^2}{\sigma_j^2}\label{defchi2}\ee where $j$ refers to the $j^{th}$ datapoint as resulted from Eq. (\ref{df}) and $dF(\alpha,\lambda,c,z_j)$ is the residual force parametrized by eq. (\ref{model}), for the same distance $z_j$ between  cantilever and microsphera, that corresponds to measured residual force $dF(j)$. Also $N$ is the number of datapoints which is $96$ for the full dataset.

\begin{figure}[b]
\centering
\vspace{0cm}\rotatebox{0}{\vspace{0cm}\hspace{0cm}\resizebox{0.49\textwidth}{!}{\includegraphics{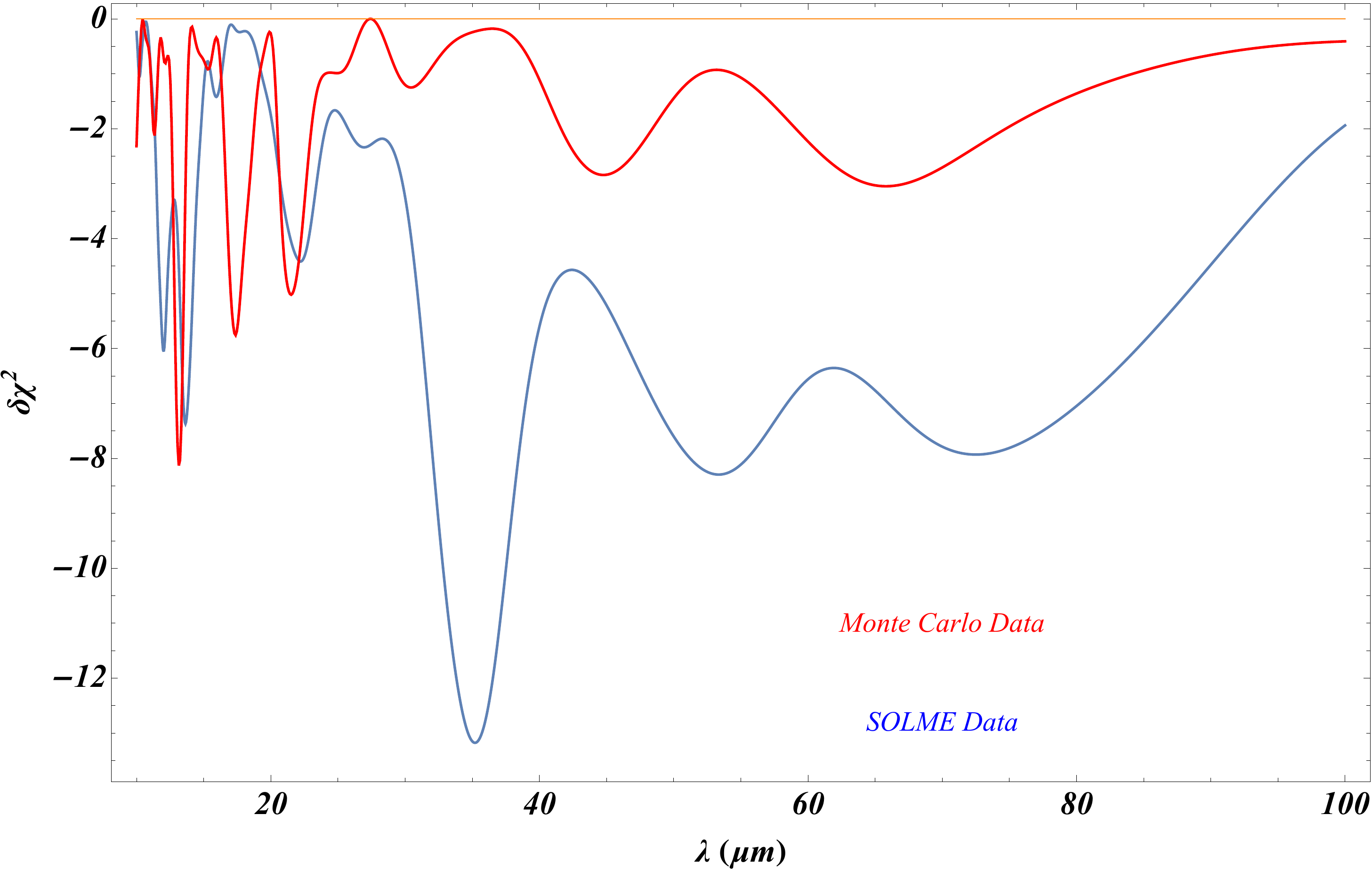}}}
\caption{The value of the minimized difference $\delta\chi^2=\chi^2_{oscillating}-\chi^2_{\alpha=0}$ as a function of the spatial wavelength $\lambda$ for the experimental data and a random Monte-Carlo dataset simulating the SOLME data under the assumption of zero residual force and gaussian errors. The depth of the $\delta\chi^2$ deepest minimum is significantly larger when the real data are fit to the oscillating parametrization.} \label{chi2allrand}
\end{figure}

\begin{figure}[b]
\centering
\vspace{0cm}\rotatebox{0}{\vspace{0cm}\hspace{0cm}\resizebox{0.49\textwidth}{!}{\includegraphics{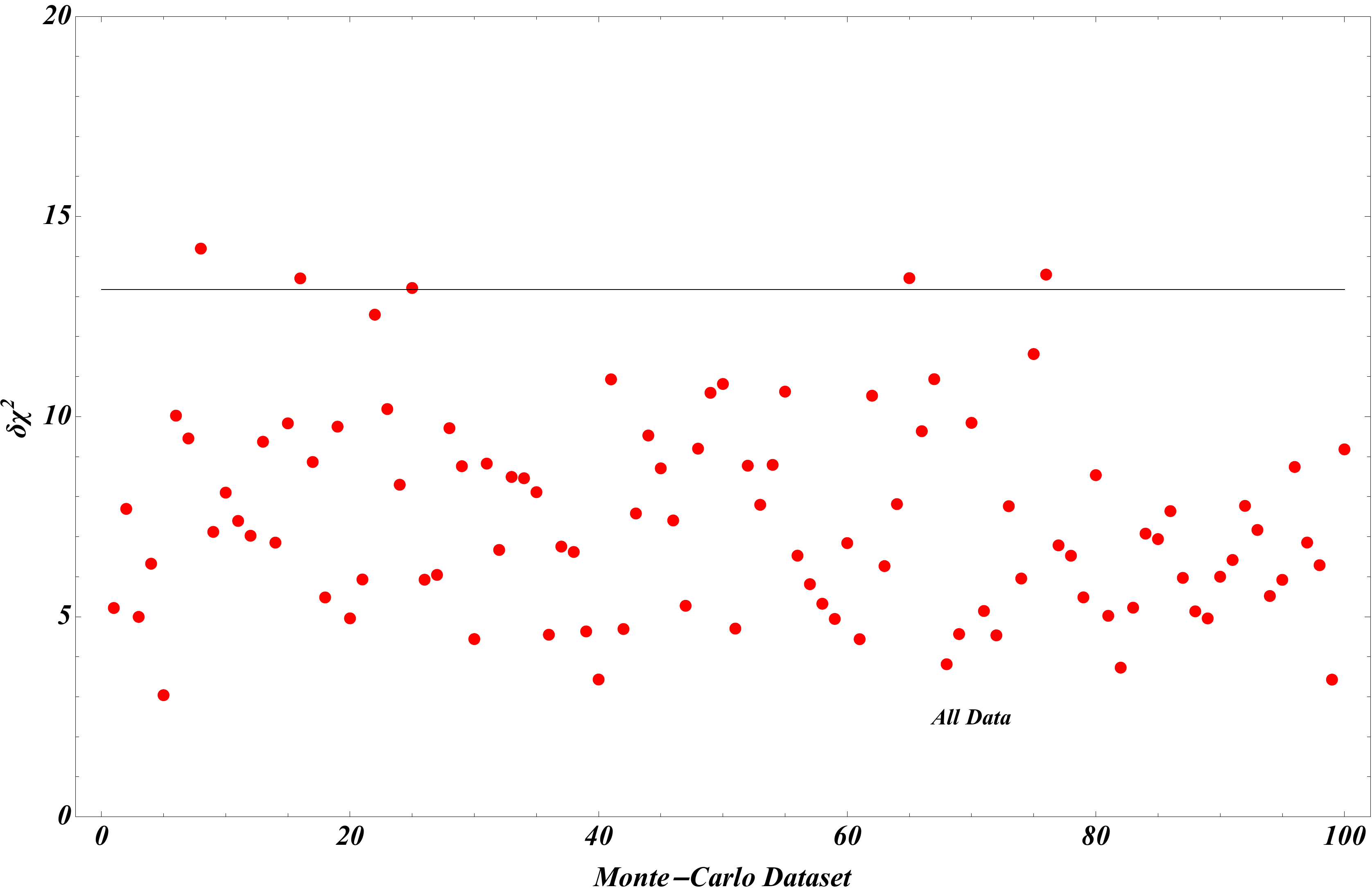}}}
\caption{\small The maximum $\delta\chi^2$ depth in the range $\lambda\in[10,100]$ for 100 Monte Carlo datasets assuming zero residual force (red dots). The horizontal line corresponds to the maximum $\delta\chi^2$ depth for the actual SOLME dataset.} \label{plall}
\end{figure}

We found that, for the full dataset, $\chi^2(\alpha,\lambda,c)$ is minimized for \ba \alpha&=&(0.011 \pm 0.004)fN \label{abf} \\ \lambda&=&(35.2\pm 0.6) \mu m \label{bbf} \\ c&=&(5.47\pm 0.06) rad\simeq 7\pi /4 \label{cbf} \ea This value of the best fit phase $c$ differs by about $\pi$ from the corresponding best fit phase obtained in Ref. \cite{Perivolaropoulos:2016ucs} when fitting the Washington experiment data to the same parametrization. The minimum value of $\chi^2$ is $\chi^2(\alpha,\lambda,c)=85.2$ compared to $\chi^2(0,\lambda,c)=98.3$ corresponding to zero residual force ($dF=0$). In Fig. \ref{chi2all} we show  the (minimized with respect to $\alpha , c$) $\chi^2(\alpha,\lambda,c)$ for the full dataset as a function of the spatial wavelength $\lambda$.  Clearly, there is a well pronounced minimum at the spatial wavelength $\lambda=35.2\mu m$.

The red horizontal line corresponds to the value of $\chi^2$ of zero residual force $dF(\alpha=0,\lambda,c,r_j)=0$. The difference between zero force residual and best fit oscillating parametrization is $\delta \chi^2=13.1$. The $1\sigma$ and $2\sigma$ contours for the two parameters $\alpha$, $\lambda$ (fixing $c=7\pi/4$) are shown in Fig. \ref{contourall}. These contours indicate that the zero residual $\alpha=0$ line is about $3\sigma$ away from the best fit $\alpha=0.01$. In Fig. \ref{bfit} we show the full dataset (residual force in $fN$ vs distance in $\mu m$) along with the best fit oscillating model  (\ref{model}). The oscillating signal in the data is clearly visible.

In view of the presence of other less deep $\chi^2$ minima at different spatial wavelengths, this $3\sigma$ estimate is an overestimate of the true significance of the oscillating signal. In order to estimate the correct statistical significance of the signal we have performed a Monte Carlo simulation. The goal of such a Monte Carlo simulation is to estimate how often would such a deep $\chi^2$ minimum occur in SOLME simulated data derived under the assumption of an underlying zero residual force.

In order to verify the level of significance of the identified oscillating signal we have generated Gaussian Monte Carlo datasets under the assumption of zero residual force. In particular, 
 we used the Normal Distribution to take random values  for the residual forces (with mean value zero) for each datapoint distance $z$ with the same standard deviation as the experimental data. We processed multiple datasets of random datapoints with the same method as the measured data. A typical form of $\delta \chi^2(\lambda)\equiv\chi^2_{oscillating}-\chi^2_{\alpha=0}$ (after minimization with respect to $\alpha$, $c$ at each value of $\lambda$) is shown in Fig. \ref{chi2allrand}. Clearly, the depth of the deepest minimum of the Monte Carlo dataset (red line) is significantly smaller than the maximum depth obtained with the real dataset (blue line).
 
We considered 100 Monte Carlo zero residual force datasets and we calculated for each Monte-Carlo dataset the deepest $\chi^2$ minimum in the range $\lambda\in[10-100] \mu m$ and subtracted this minimum $\chi^2$ from the corresponding of zero residual force $\chi^2$ obtained from the Monte-Carlo data. 
Thus we calculated the difference
\be \delta\chi^2= \chi^2_{zero\; residual}-\chi^2_{min-oscillating\;residual}\label{dchi}\ee
For the real data this corresponds to the difference $\delta\chi^2$ between the $\alpha=0$ red line of Fig. \ref{chi2all} and the deepest minimum of the blue line. This is the horizontal line in Fig. \ref{plall} at $\delta \chi^2=13.1$. For the Monte Carlo datasets this difference corresponds to the difference between the deepest minimum of the red line and the horizontal red line of Fig. \ref{chi2allrand}. Each one of the red dots of Fig. \ref{plall} corresponds to such Monte Carlo difference. Clearly if all the $100$ red dots were found below the horizontal line of Fig. \ref{plall} ($\delta\chi^2<13.1$) then there would be less than $1\%$ probability that the deep $\chi^2$ minimum of Fig. \ref{chi2all} is due to a statistical fluctuation. Instead we find that about $5\%$ of the zero residual simulated data lead to deeper $\chi^2$ minima (five red dots in Fig. \ref{plall} are above the horizontal line). Thus the true level of significance of the oscillating signal is at about $2\sigma$. A similar effect leading to reduced level of significance compared to the one indicated by the $\chi^2$ contour plot was observed and discussed in Ref. \cite{Perivolaropoulos:2016ucs}.

We conclude that there is evidence for an oscillating signal at the $2\sigma$ level in the SOLME data. Since there is only about $5\%$ probability that this signal is due to a statistical fluctuation, most likely it is due either to a systematic effect that was not discussed in Ref. \cite{Rider:2016xaq} or it is due to new physics. Private communication with the authors of Ref. \cite{Rider:2016xaq} has indicated that the signal is most probably due to a systematic effect caused by a background due to non-Gaussian tails of the laser beam whose pressure levitates the microsphere.  Due to diffraction, the intensity of these non-Gaussian tails has a periodic oscillation, which can mimic a spatially oscillating force signal. Thus the amplitude of this detected signal can only be useful as an upper bound to the amplitude of new spatially oscillating forces on sub-mm scales.

In addition to the oscillating parametrization (\ref{model}) we have tried to fit the data using various monotonic parametrizations like a Yukawa parametrization of the form
\be dF(\alpha,\lambda,z)=\alpha\;  e^{z/\lambda}\label{modelyuk}\ee
However, in all cases the improvement of the quality of fit was minor with $\delta \chi^2 < 1$ and thus we will not discuss these cases further in this section. 

The oscillating parametrization (\ref{model}) is a phenomenological parametrization which can not be used as is to impose constraints on fundamental parameters. In order to impose such constraints the macroscopic residual force parametrization must be derived starting from a fundamental theory. For example we may assume a gravitational origin of the signal and derive the macroscopically induced residual force starting from a modified Newtonian potential of the form (\ref{vosc}). Thus we may derive the predicted macroscopic residual force between cantilever and microsphere in terms of the fundamental parameters $\alpha_O$ and $\lambda$ of eq. (\ref{vosc}) by evaluating the source integral \ref{potsourceint})  over the cantilever. This derived effective residual force may then be fit to the SOLME data leading to constraints on the fundamental parameters $\alpha_O$ and $\lambda$ rather than the corresponding phenomenological parameters of eq. (\ref{model}). This task is undertaken in the next section.

\section{Constraints on Fundamental Parameters: Source Integral}
\label{sec:sourceint}

\subsection{Newtonian Force between a Cylindrical Cantilever and a Microsphere}

\begin{figure}[b]
\centering
\vspace{0cm}\rotatebox{0}{\vspace{0cm}\hspace{0cm}\resizebox{0.5\textwidth}{!}{\includegraphics[clip, trim=0.0cm 18cm 3cm 1cm, width=1.00\textwidth]{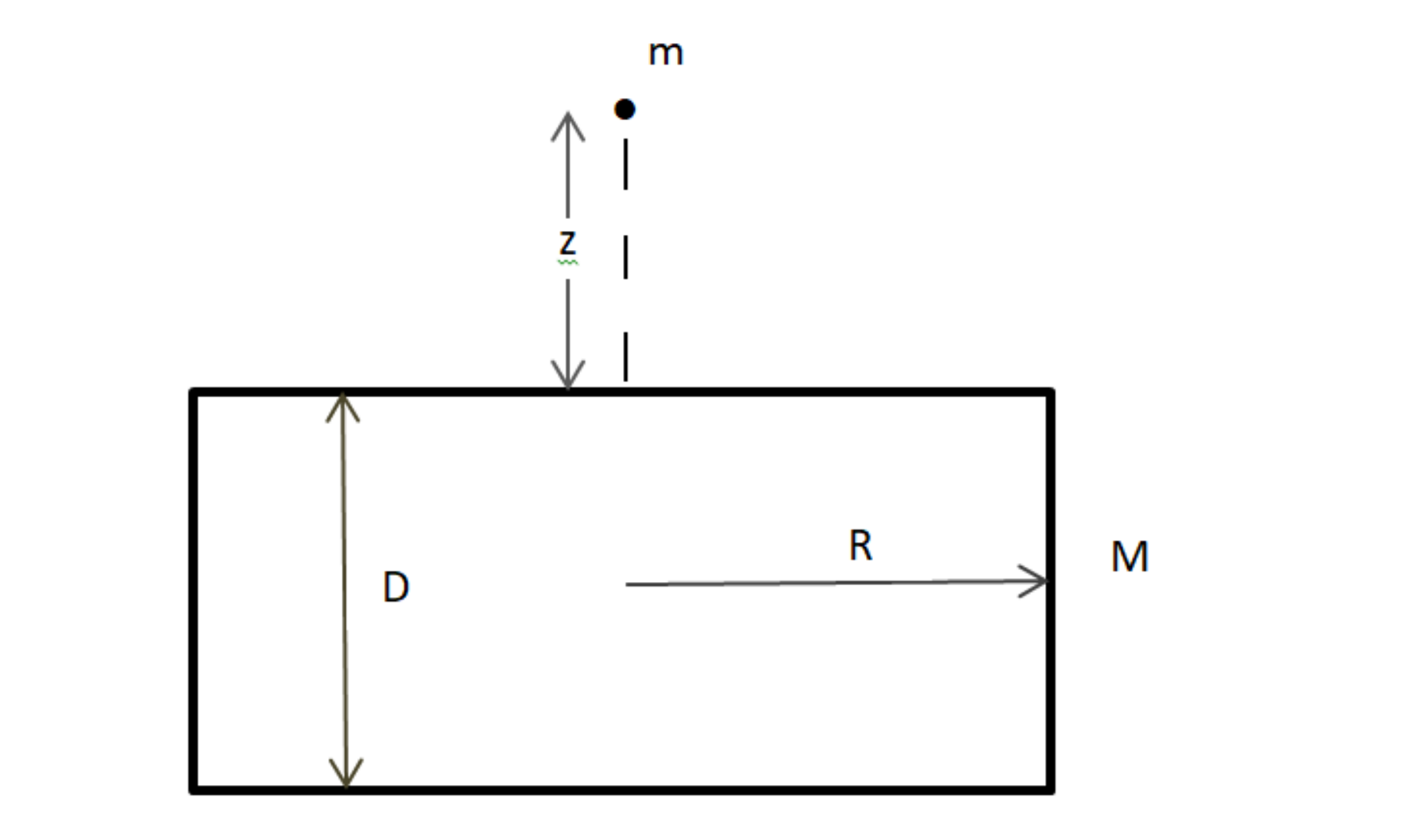}}}
\caption{\small The cantilever approximated as a cylinder and a point mass $m$ at distance $z$ from its surface. } \label{cynindgeom}
\end{figure}

We approximate the orthogonal cantilever of the SOLME by a cylindrical one of the same base area and height as the one used in the experiment. This allows for analytical evaluation of the source integral and of the macroscopic gravitational forces of the cantilever on the small microsphere located at a distance $z$ along the symmetry axis from the center of the base of the cylindrical cantilever.  Such a cantilever would have a radius $R\simeq 40\mu m$, height $D=2000\mu m$ (see Fig. \ref{cynindgeom}) and density $\rho=2.3gr/cm^3$. As stated in the previous section the mass of the microsphere was $m=0.13ng$ and its radius $r=2.5\mu m$.

We first calculate the Newtonian gravitational force between this cantilever and the microsphere. The gravitational potential energy between the cantilever and a point mass $m$ at distance $z$ from its surface is of the form 

\be V_N(z)=-2\pi \rho Gm \int_0^R rdr\int_z^{z+D}\frac{dz'}{\sqrt{r^2+z'^2}}\label{newtmacrpot}\ee

We now introduce a rescaling to dimensionless length dividing all lengths by the cantilever radius $R$ and denote with a `bar' the new dimensionless quantities. Under this rescaling the potential (\ref{newtmacrpot}) takes the form

 \begin{equation}
 V_N(\bar{z})=- \underbrace{2\pi \rho GmR^2}_{ V_1} \underbrace{\int_0^1 r'dr'\int_{\bar{z}}^{\bar{z}+\bar{D}} \frac{dz''}{\sqrt{r'^2+z''^2}}}_{\bar{V}_N(\bar{z})}\label{dimlessnewtv}
\end{equation}
where the definitions of the potential unit $V_1$ and of the dimensionless gravitational potential $\bar{V}_N$ are shown in eq. (\ref{dimlessnewtv}).
The corresponding $z$ component of the interaction force is 
\begin{equation}
F_{zN}(\bar{z})=- \underbrace{ 2\pi Gm\rho R}_{F_1}  \underbrace{ \frac{\partial \bar{V}_N(\bar{z})}{\partial \bar{z}}}_{\bar{F}_{zN}(\bar{z})}\label{newtf1}
\end{equation} 
It is straightforward to calculate the dimensionless part of the force $\bar{F}_{zN}(\bar{z})\equiv \frac{\partial \bar{V}_N(\bar{z})}{\partial \bar{z}}$ as 

\be \bar{F}_{zN}(\bar{z},\bar{D})=-\bar{D}-\sqrt{1+\bar{z}^2}+\sqrt{1+(\bar{D}+\bar{z})^2}\label{dimlessnewtf}\ee 

For small $\bar{z}$ ($\bar{z}\ll 1 $) this is a constant as expected 
\be  \bar{F}_{zN}(\bar{z},\bar{D})\simeq \sqrt{1+\bar{D}}-(1+\bar{D}) \label{smallznewtf}\ee 

\noindent while for $\bar{z}\gg 1$ it also has the anticipated asymptotic behavior as an inverse square of the distance \be \bar{F}_{zN}(\bar{z},\bar{D})\simeq -\frac{\bar{D}}{2\bar{z}^2}\label{largeznewtf}\ee

The dimensions corresponding to the SOLME are  $R=40\mu m$, $\bar{D}=50$, $\bar{r}_1=0.0625$, $\bar{z}_{min}=0.5$, $\bar{z}_{max}=6.25$ where $\bar{r}_1\equiv \frac{r_1}{R}$ is the dimensionless form of the radius of the microsphere which is clearly much smaller than all the other dimensions of the experiment.  In view of this fact we may approximate the microsphere as a point mass and assume that the predicted Newtonian force on it is provided to a good approximation by eqs. (\ref{newtf1}) and (\ref{dimlessnewtf}).

An improved approximation for the calculation of the Newtonian force on the microsphere is the averaging of the force through the evaluation of the integral 

\be \bar{F}_{zN,total}(\bar{z},\bar{D},\bar{r}_1)=\frac{1}{2\bar{r}_1}\int_{{\bar z}_0-\bar{r}_1}^{{\bar z}_0+\bar{r}_1}dz'\bar{F}_{zN}(z',\bar{D})\label{newtforceav}\ee

We have found that this improved approximation has a minor effect (less than $1\%$) on the estimated force on the micropshere. Thus, in what follows we approximate the micropshere as a point mass that is subject to a Newtonian force from the cantilever provided by eqs. (\ref{newtf1}) and (\ref{dimlessnewtf}) as
\be 
F_{zN}(\bar{z},\bar{D})=\alpha_N\times \underbrace{ 2\pi Gm\rho R}_{F_1}\times \bar{F}_{zN}(\frac{z}{40},50) \label{newtforcefin}\ee
where $z$ is in $\mu m$ and  $F_1=2\pi Gm\rho R \simeq 5\times 10^{-9} fN$ for the geometry and objects used in the SOLME. We have allowed for a short range amplification factor $\alpha_N$ to the Newtonian force. 
\begin{figure}[t]
\centering
\vspace{0cm}\rotatebox{0}{\vspace{0cm}\hspace{0cm}\resizebox{0.49\textwidth}{!}{\includegraphics{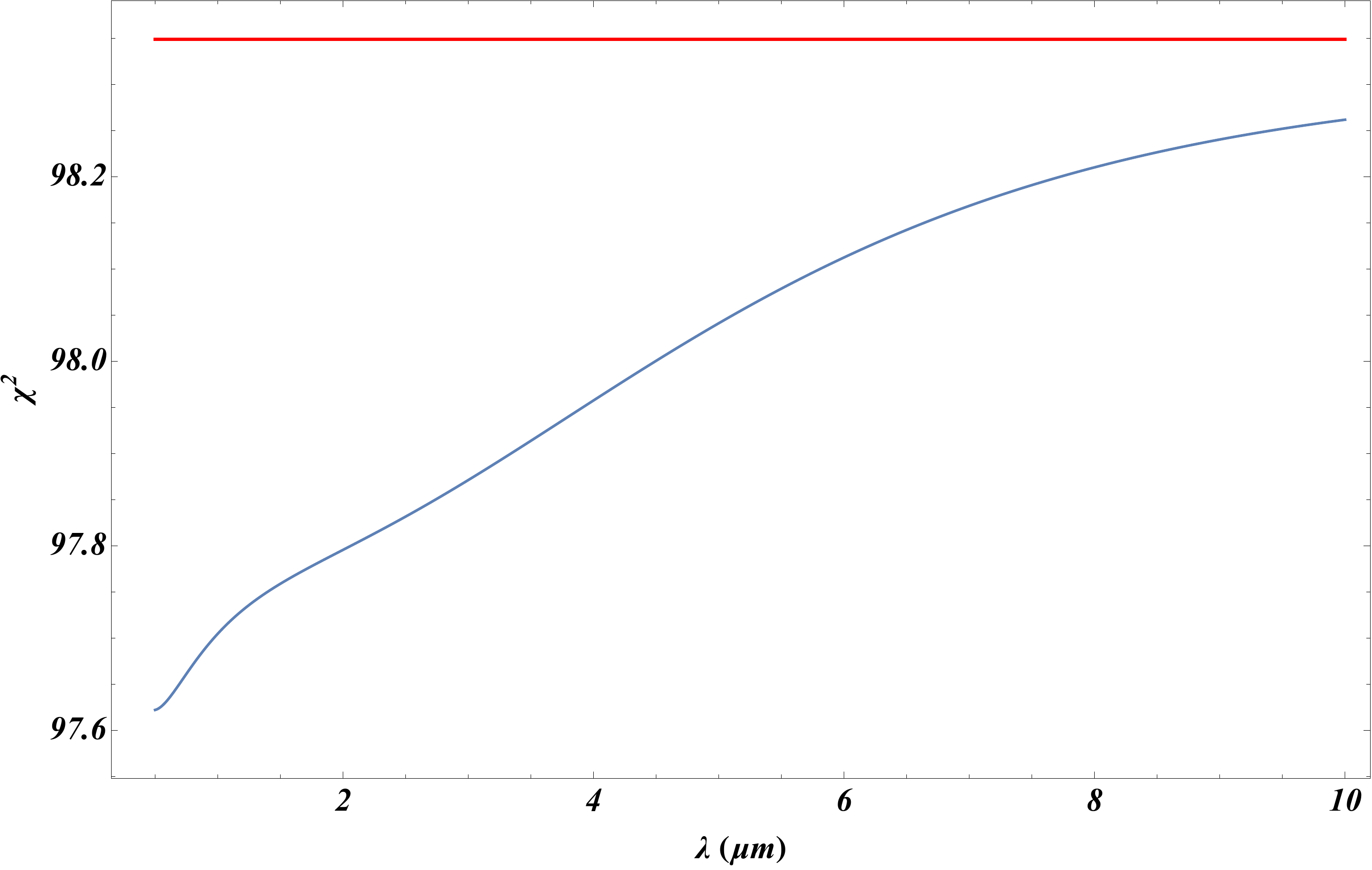}}}
\caption{The minimized $\chi^2$ using the SOLME data as a function of the parameter $\lambda$ of the Yukawa force ansatz including the effects of the source integral. The impovement of the fit is marginal despite the additional two parameters $\alpha_Y, \lambda$.} \label{chi2yuk}
\end{figure}
Since $\bar{F}_{zN}(\frac{z}{40},50)<1$ for the distances considered in the SOLME ($z>20\mu m$) it is clear that the Newtonian force is much smaller than the residual forces measured in the SOLME which are of $O(10^{-2}) fN$ and an amplification by a factor  $\alpha_N\simeq 10^7$ on these scales would be required for such a force to be observable by the SOLME. 
\begin{figure}[t]
\centering
\vspace{0cm}\rotatebox{0}{\vspace{0cm}\hspace{0cm}\resizebox{0.49\textwidth}{!}{\includegraphics{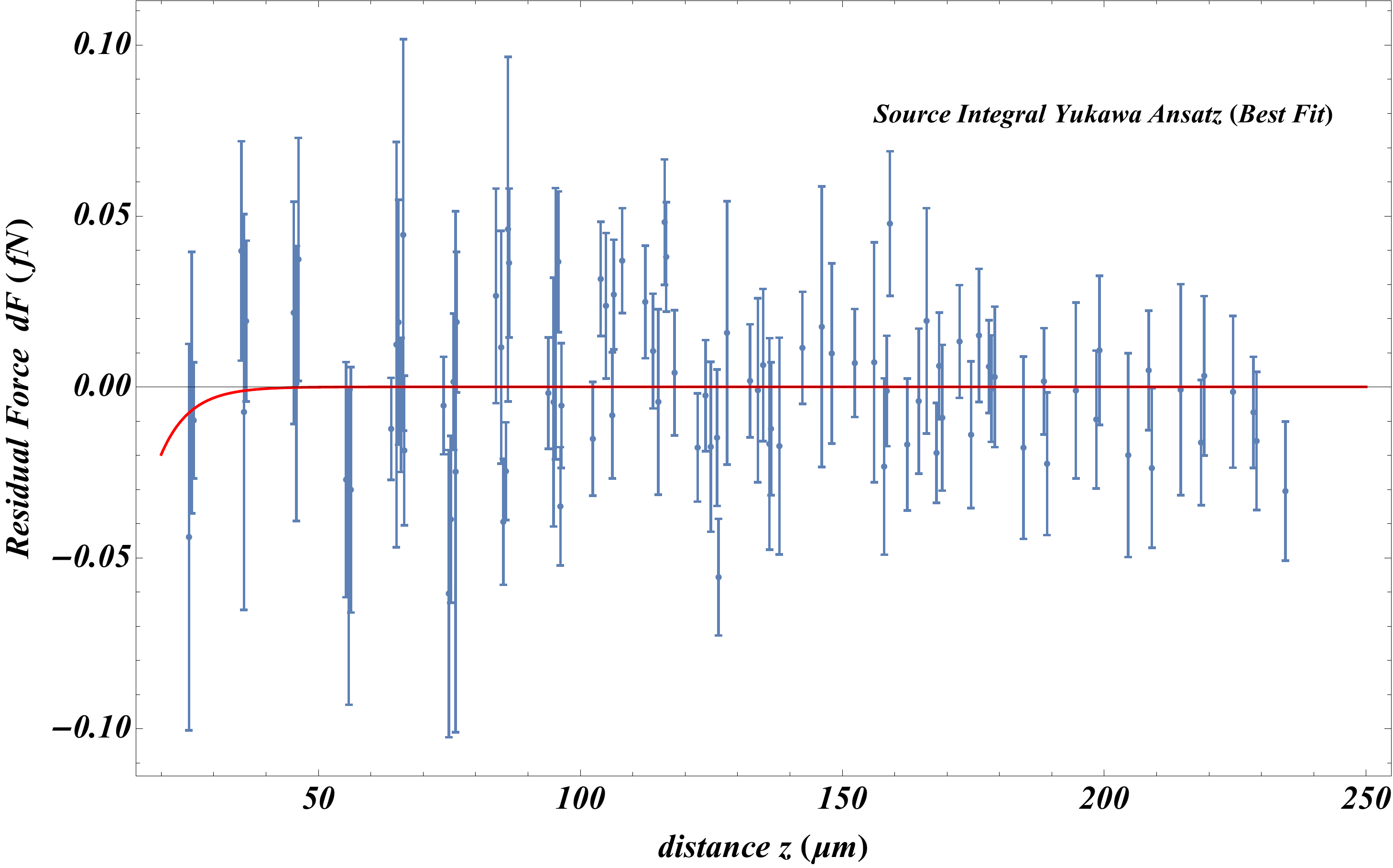}}}
\caption{The best fit form of the the source integral for the Yukawa residual force is practically indistinguishable from the zero force residual.} \label{bfityuk}
\end{figure}

\subsection{Yukawa and Power Law Residual Force between a Cylindrical Cantilever and a Microsphere}

Deviations from the Newtonian potential on sub-millimeter scales can be parameterized through a Yukawa interaction, an oscillating model or a power law parametrization. In the case of the Yukawa deviation, the potential energy of a point mass $M$ interacting with a point mass $m$ at a distance $r$ gets generalized as $V(r)=V_N(r)+V_Y(r)$ with 
\be V_Y(r)=-\frac{GMm}{r}\alpha_Y e^{-\frac{r}{\lambda}}\label{yukpot1}\ee 
where $\alpha_Y$ and $\lambda$ are appropriate parameters to be constrained.
In this case, the Yukawa rescaled dimensionless interaction potential energy (see eq. (\ref{newtmacrpot}))  between a cylinder of dimensionless height $\bar{D}$ (the cantilever) and a point mass $m$ (the microsphere) located at a distance $\bar{z}$ from the center of one of the cylinder bases is
\be \bar{V}_Y(\bar{z})=-\alpha_Y \int_0^1 r'dr'\int_{\bar{z}}^{\bar{z}+\bar{D}}dz'\frac{e^{-\frac{\sqrt{r'^2+z'^2}}{\bar{\lambda}}}}{\sqrt{r'^2+z'^2}}\label{yukmacrpot}\ee 

The corresponding $z$ component of the force $\bar{F}_{zY}(\bar{z})\equiv \frac{\partial \bar{V}_Y(\bar{z})}{\partial \bar{z}}$ induced on the mass $m$ can be analytically evaluated by first obtaining the source integral (\ref{yukmacrpot}) The result is
\begin{widetext}
\be \bar{F}_{zY}\left(\bar{z},\bar{D},\bar{\lambda}\right)=\alpha_Y \lambda \left(e^{-\frac{\bar{D}+\bar{z}}{\bar{\lambda}}}+ e^{-\frac{\sqrt{1+\bar{z}^2}}{\bar{\lambda}}}-e^{-\frac{\bar{z}}{\bar{\lambda}}}-e^{-\frac{\sqrt{1+(\bar{D}+\bar{z})^2}}{\bar{\lambda}}}\right)\label{yukmacrforce1}\ee 
\end{widetext}
with $\bar{\lambda}=\frac{\lambda}{R}$. The asymptotic behaviour of the macroscopic Yukawa force is as expected namely it is exponentially suppressed for $\bar{z}>>1$ while for small $\bar{z}$ it is constant approximated as
\be \bar{F}_{zY}(\bar{z},\bar{D},\bar{\lambda})=\alpha_Y \bar{\lambda}(-1+e^{-\frac{1}{\bar{\lambda}}}+e^{-\frac{\bar{D}}{\bar{\lambda}}}-e^{-\frac{\sqrt{1+\bar{D}^2}}{\bar{\lambda}}})\label{yukforcelargez}\ee
For the SOLME the full residual Yukawa force may be expressed as 
\be F_{zY,tot}=\bar{F}_{zY,tot}(\frac{z}{40},50,\frac{\lambda}{40})\times 5\times \underbrace{ 10^{-9}\times \alpha_Y}_{\alpha_{Y9}}\label{solmeyukforce}\ee where $z$, $\lambda$ must be substituted in $\mu m$ and the force is in $fN$. We have found that as in the case of the simple phenomenological Yukawa parametrization discussed in the previous section, the source integral Yukawa force (\ref{solmeyukforce}) is unable to improve the fit of the SOLME residual force data by more than 1 ($\delta\chi^2<1$) compared to the zero residual force parametrization. This is demonstrated  in Fig. \ref{chi2yuk} where we show the minimum value of $\chi^2$ as a function of $\lambda$ for the macroscopic Yukawa force residual (\ref{solmeyukforce}) and for the zero force residual (red line). Clearly we have $\delta \chi^2<1$ for all values of $\lambda$ considered. Thus the Yukawa potential does not provide a more efficient macroscopic residual force parametrization for fitting the force residuals of the SOLME data compared to null hypothesis of the zero force residual. This is also demonstrated in Fig. \ref{bfityuk} where we show the best fit Yukawa residual force for $\alpha_Y=1$ which is achieved for $\lambda=5.6 \mu m$ and is practically indistinguishable from the zero residual force for most of the range of the force residual SOLME data. 

A similar conclusion is obtained for other monotonic residual force parametrizations like power law deviations from the Newtonian potential. In that case the generalized gravitational potential would be of the form  $V(r)=V_N(r)+V_P(r)$ with 
\be V_P(r)=-\frac{\alpha_P GMm}{r^n}\label{powerlawpot}\ee 
The rescaled dimensionless source integral may be written as 
\be \bar{V}_P(\bar{z})=-\alpha_P \int_0^1 r'dr'\int_{\bar{z}}^{\bar{z}+\bar{D}}\frac{dz'}{(r'^2+z'^2)^{n/2}}\label{powlawsint}\ee leading to the $z$ component of the rescaled dimensionless force  $\bar{F}_{zP}(\bar{z})\equiv \frac{\partial \bar{V}_P(\bar{z})}{\partial \bar{z}}$ in the analytic form 
\begin{widetext}
\begin{eqnarray}
 &&\bar{F}_{zP}(\bar{z},\bar{D},n)=\alpha_P\frac{\bar{z}^{-n}(\bar{D}+\bar{z})^{-n}}{n-2}\bigg[(1+\bar{z}^2)\left[1+(\bar{D}+\bar{z})^2\right]\bigg]^{\frac{-n}{2}}
  \Bigg\{-\bar{z}^n(\bar{D}+\bar{z})^n(1+\bar{z}^2)^\frac{n}{2}\left[1+(\bar{D}+\bar{z})^2\right]\nonumber\\&&+\left[1+(\bar{D}+\bar{z})^2\right]^\frac{n}{2}\bigg[\bar{z}^n
 (\bar{D}+\bar{z})^n(1+\bar{z}^2)+(1+\bar{z}^2)^\frac{n}{2}\left[\bar{z}^n(\bar{D}+\bar{z})^2-\bar{z}^2(\bar{D}+\bar{z})^n\right]\bigg] \Bigg\}\label{powerlawforce}
 \end{eqnarray}
\end{widetext}

Introducing the parameters of the SOLME the dimensionful force in $fN$ takes the form  \be F_{zP,tot}=\bar{F}_{zP,tot}(\frac{z}{40},50,n)\times 5\times \underbrace{ 10^{-9}\times \alpha_P}_{\alpha_{Pg}}\label{eq28}\ee

It is straightforward to show that the quality of fit of this power law source integral force residual is similar to that of the coresponding Yukawa residual and thus it is not of particular interest since it is not favoured over the zero residual hypothesis. Thus we will not pursue this case further.

\subsection{Oscillating Force Residual between a Cylindrical Cantilever and a Microsphere}

We now consider an oscillating gravitational residual potential of the form  $V(r)=V_N(r)+V_O(r)$ with \be V_O(r)=-\frac{GMm}{r}\alpha_O \cos\left(\frac{2\pi}{\lambda}r+\theta\right)\label{eq18}\ee

The macroscopic dimensionless form of the potential energy between a cylindrical cantilever and a microsphere on the cantilever's  axis of symmetry is expressed in terms of the dimensionless source integral as

\be \bar{V}_O(\bar{z})=-\alpha_O \int_0^1 r'dr'\int_{\bar{z}}^{\bar{z}+\bar{D}}dz'\frac{\cos\left({\frac{2\pi \sqrt{r'^2+z'^2}}{\lambda}}+\theta\right)}{\sqrt{r'^2+z'^2}}\label{eq19}\ee 
\begin{figure}[t]
\centering
\vspace{0cm}\rotatebox{0}{\vspace{0cm}\hspace{0cm}\resizebox{0.49\textwidth}{!}{\includegraphics{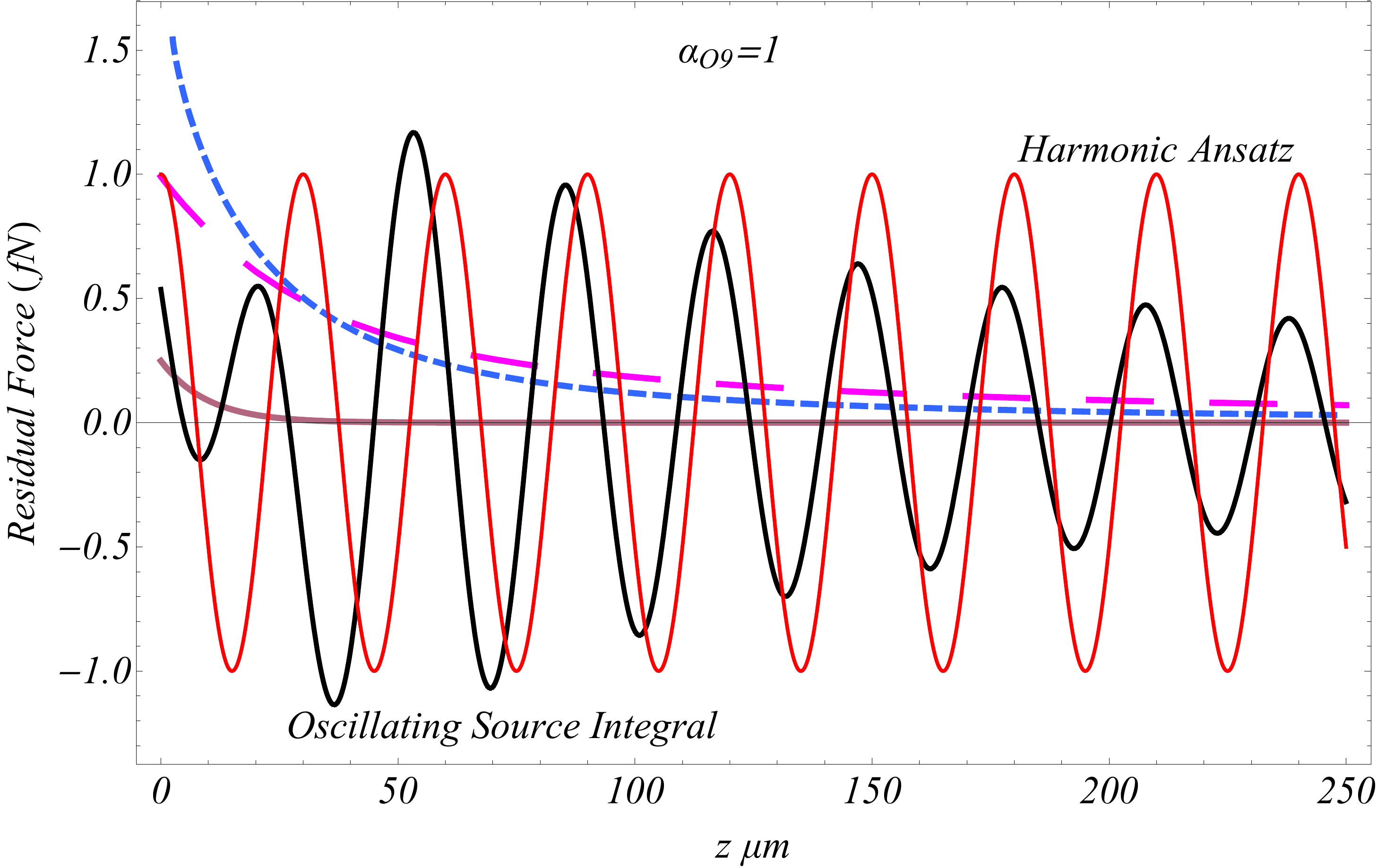}}}
\caption{Comparison between the  source  integral signal for oscillating force with the naive constant amplitude cosine oscillator. On scales $z$ larger than the disk radius ($R=40\mu m$) they both behave like harmonic functions with very similar wavelength,while on small scales the signal is not periodic. Also, we have plot the Newtonian ansatz (blue line), the Yukawa ansatz (gray line) and a power law force with $n=1.5$ (magenta line). For the oscillating source integrals we have set $\alpha_{O9}=1$ which implies $\alpha_O=10^9$.} \label{plforces}
\end{figure}
 The corresponding $z$ component of the rescaled dimensionless force $\bar{F}_{zO}(\bar{z})\equiv \frac{\partial \bar{V}_O(\bar{z})}{\partial \bar{z}}$, can be obtained analytically as
\begin{widetext}
\be
\bar{F}_{zO}(\bar{z},\bar{D},\bar{\lambda},\theta)=\frac{\alpha_O {\bar{\lambda}}}{2\pi}\left[\sin\left(\frac{2\pi \bar{z}}{\bar{\lambda}}+\theta\right)-\sin\left(\frac{2\pi(\bar{D}+\bar{z})}{\bar{\lambda}}+\theta\right)+ \sin\left(\frac{2\pi \sqrt{1+(\bar{D}+\bar{z})^2}}{\bar{\lambda}}+\theta\right)-\sin\left(\frac{2\pi \sqrt{1+\bar{z}^2}}{\bar{\lambda}}+\theta\right)\right]\label{oscilmacroforce1}
 \ee
\end{widetext}
\begin{figure*}[ht] \centering
\begin{center}
$\begin{array}{@{\hspace{-0.10in}}c@{\hspace{0.0in}}c}
\multicolumn{1}{l}{\mbox{}} & \multicolumn{1}{l}{\mbox{}} \\
[-0.2in] \epsfxsize=3.4in \epsffile{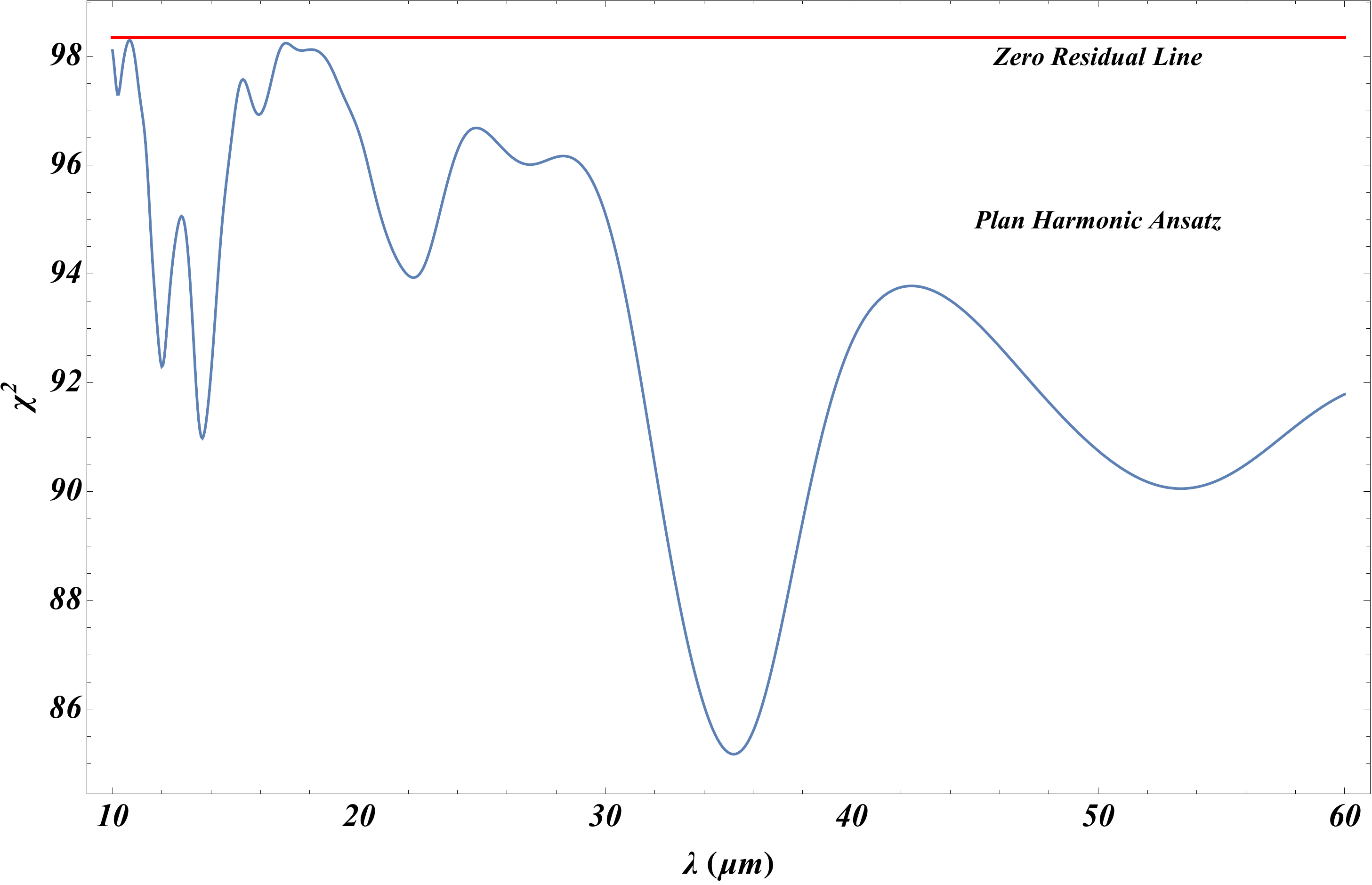} & \epsfxsize=3.4in
\epsffile{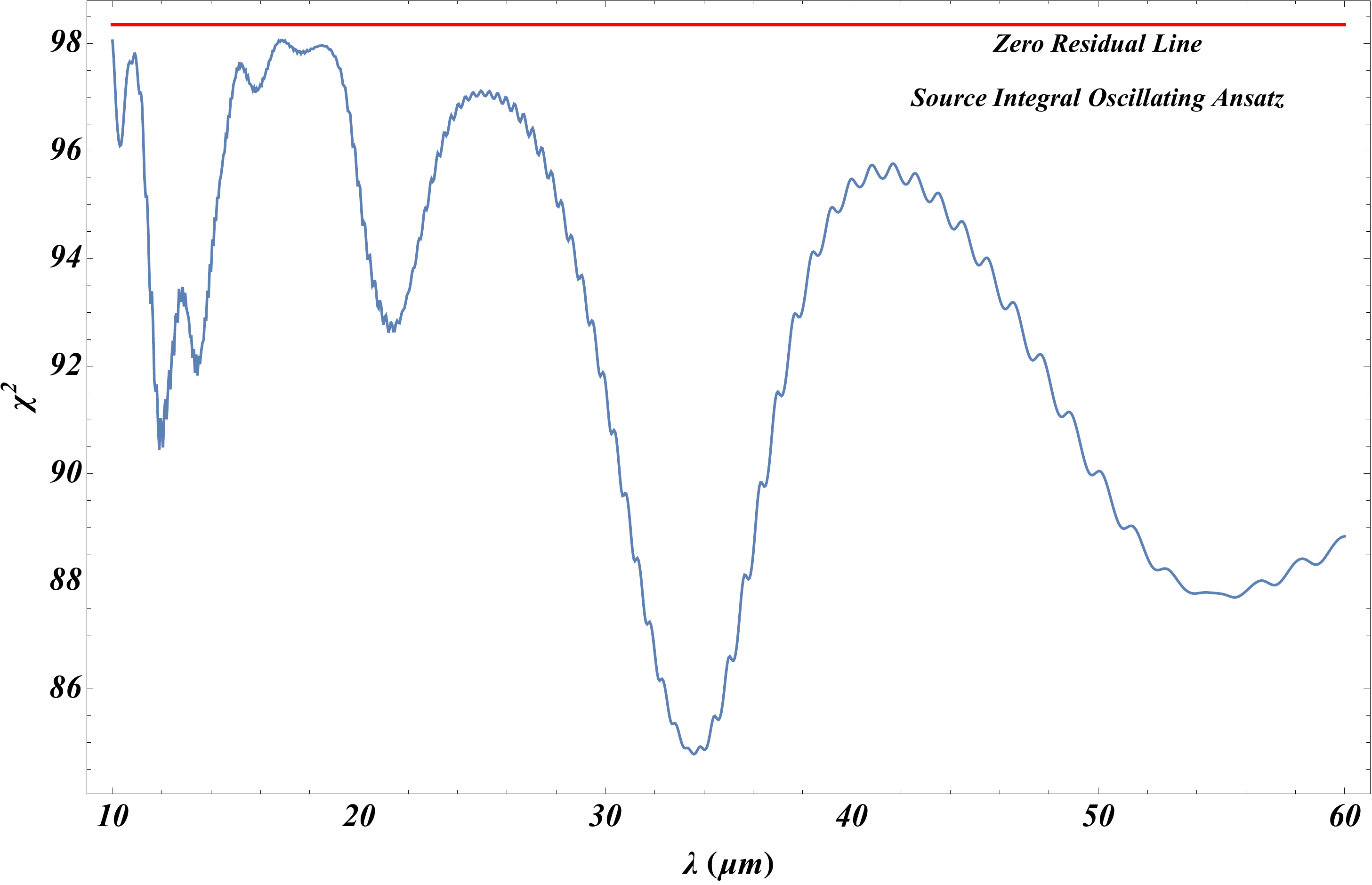} \\
\end{array}$
\end{center}
\vspace{0.0cm} \caption{\small Left panel: The $\chi^2$ value as a function of the wavelength $\lambda$ for plain harmonic ansatz.  Right panel: The $\chi^2$ value for the source integral oscillating ansatz. In both cases the fit improvement to the data is significant compared to a null residual fit. The difference in $\chi^2$ is more than 13 units and the minimum appears in almost the same wavelength (about $35\mu m$).} \label{chiosc}
\end{figure*}
For large $z$, the residual force (\ref{oscilmacroforce1}) is oscillating with an amplitude that decreases as $1/\bar{z}$ and is of the form  
\begin{widetext}
\be
\bar{F}_{zO}\left(\bar{z},\bar{D},\bar{\lambda},\theta\right)=\cos{\left(\frac{2\pi \bar{z}}{\bar{\lambda}}\right)}\frac{\cos{\left(\frac{2\pi \bar{D}}{\bar{\lambda}}+\theta\right)}-\cos{\theta}}{2\bar{z}}-\sin{\left(\frac{2\pi \bar{z}}{\bar{\lambda}}\right)}\frac{\sin{\left(\frac{2\pi \bar{D}}{\bar{\lambda}}+\theta\right)}-\sin{\theta}}{2\bar{z}}
\label{oscilmacroforcelargez}
\ee
For small $\bar{z}$ we find 
\be
 \bar F_{zO}\left(\bar{z},\bar{D},\bar{\lambda},\theta\right)=\frac{\bar{\lambda}}{2\pi}\left[\sin{\theta}-\sin{\left(\frac{2\pi}{\bar{\lambda}}+\theta\right)}-\sin{\left(\frac{2\pi \bar{D}}{\bar{\lambda}}+\theta
\right)}+\sin{\left(\frac{2\pi \sqrt{1+\bar{D}^2}}{\bar{\lambda}}+\theta\right)}\right]\label{oscilmacroforcz0}
\ee
\end{widetext}
\begin{figure*}[ht] \centering
\begin{center}
$\begin{array}{@{\hspace{-0.10in}}c@{\hspace{0.0in}}c}
\multicolumn{1}{l}{\mbox{}} & \multicolumn{1}{l}{\mbox{}} \\
[-0.2in] \epsfxsize=3.4in \epsffile{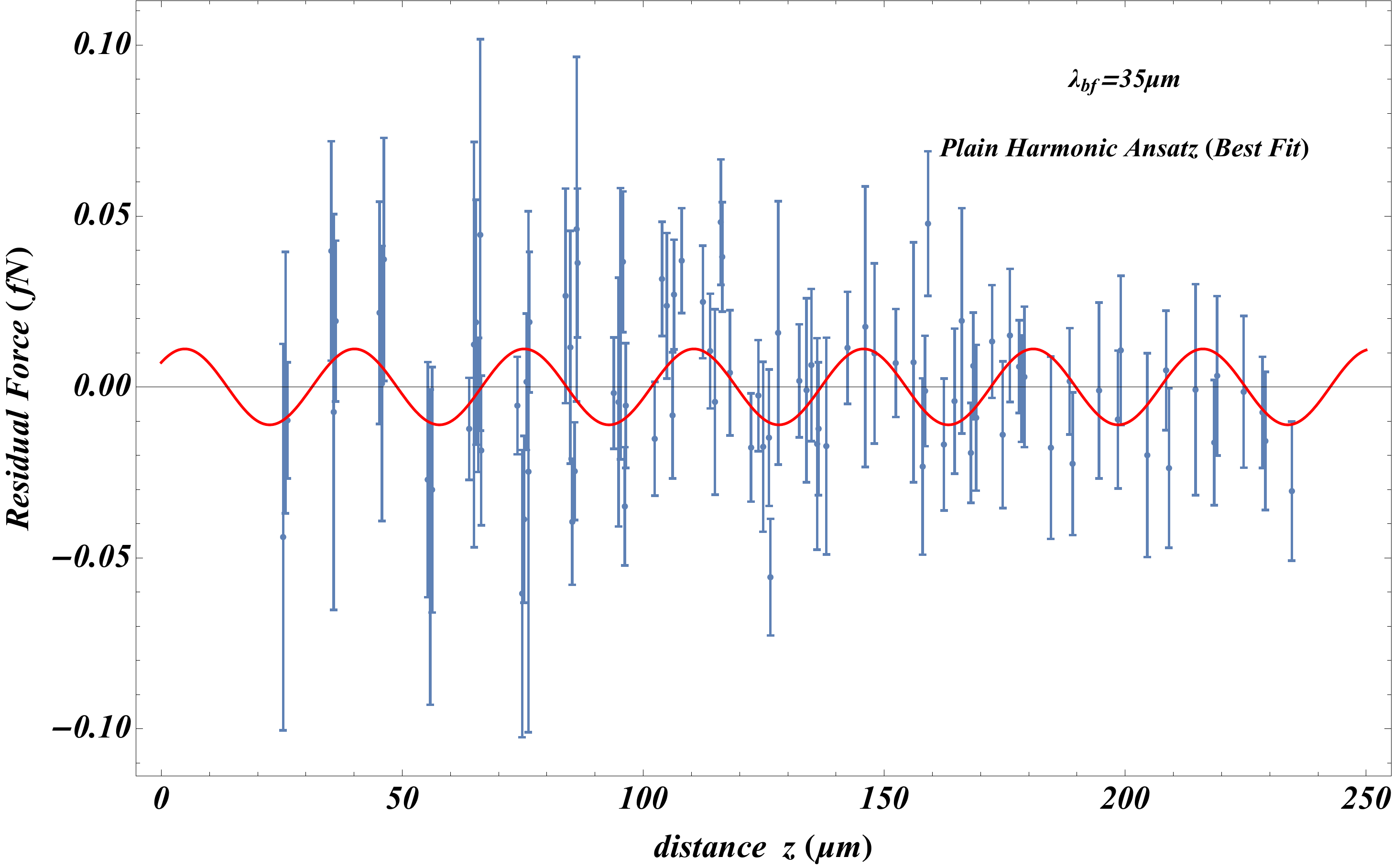} & \epsfxsize=3.4in
\epsffile{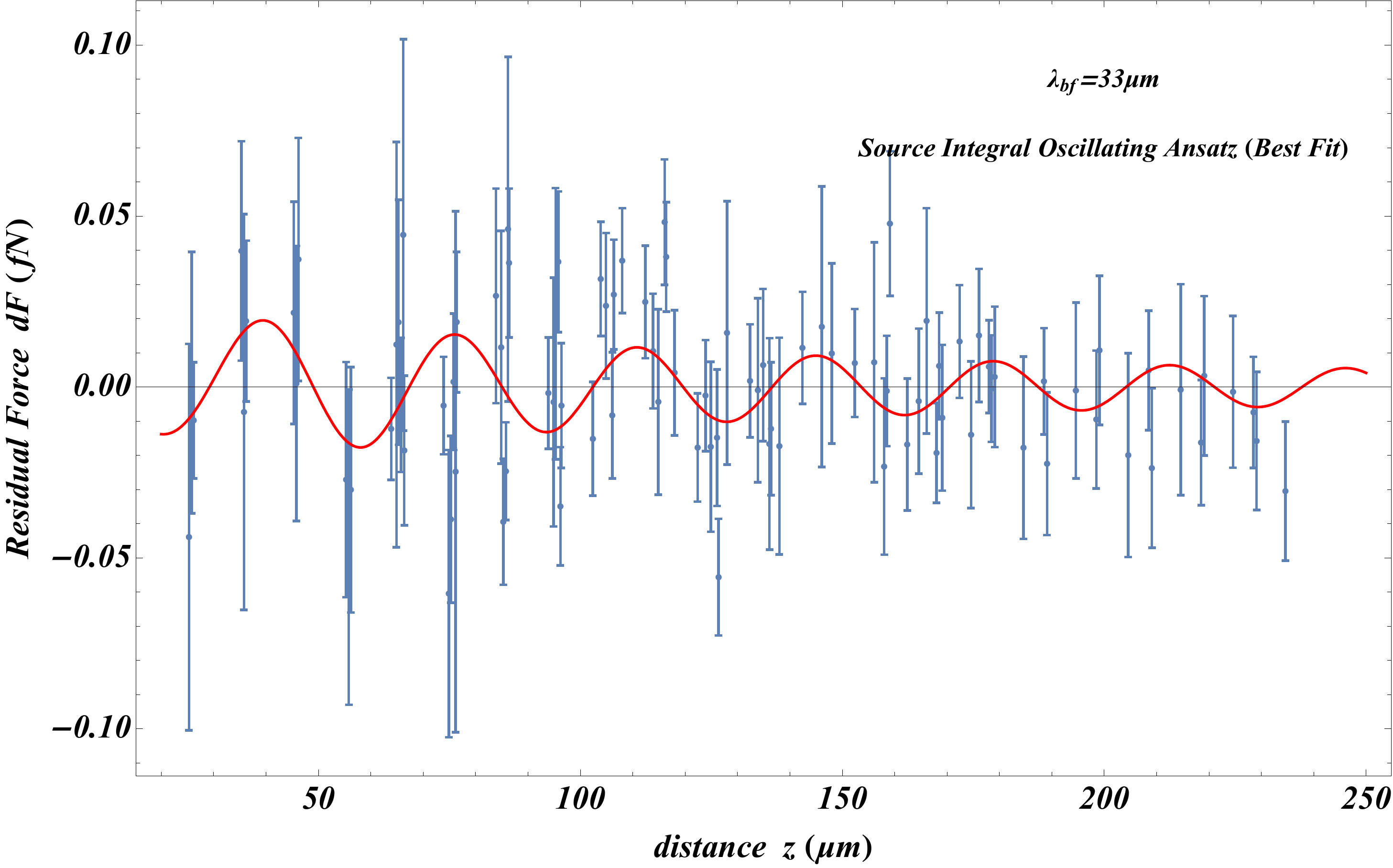} \\
\end{array}$
\end{center}
\vspace{0.0cm} \caption{\small Left panel: The best fit plain harmonic ansatz with $\lambda=35\mu m$. Right panel: The best fit source integral oscillating ansatz with $\lambda=33\mu m$.  As we see, in both cases the waveform is practically the same even though the amplitude for the best fit source integral decreases slowly with distance.} \label{bfitnew}
\end{figure*}
For the case of the SOLME the oscillating residual force on the microsphere located at a distance $z\mu m$ from the cantilever is would be of the form

\be F_{zO,tot}=\bar{F}_{zO,tot}(\frac{z}{40},50,\frac{\lambda}{40},\theta)\times 5\times \underbrace{ 10^{-9}\times \alpha_O}_{\alpha_{O9}}\label{solmeoscilforce}\ee where $z$, $\lambda$ in $\mu m$ and the force is  in $fN$.

In Fig. \ref{plforces} the force source integral (\ref{solmeoscilforce}) with $\lambda=30\mu m$ and $\alpha_{O9}=1$ (thick black dotted line) is compared with the plain harmonic residual force (\ref{model}) with the same $\lambda$ and $\alpha=1$ (continous red line), with the Newtonian source integral force (\ref{dimlessnewtf}) (long dashed line), with a power law source integral ($n=1.5$, (\ref{powerlawforce}), blue dashed line) and with a Yukawa source integral force with $\lambda=10\mu m$ (gray line). Notice that the oscillating force source integral for the particular parameters is an oscillating non-periodic function with initially increasing amplitude which reaches a maximum and subsequently decreases at large distances in accordance with the predicted asymptotic behaviour (\ref{oscilmacroforcelargez}).

It is straightforward to fit the SOLME data using the macroscopic oscillating residual force (\ref{solmeoscilforce}) obtained from the source integral. In this case as in the case of the plain harmonic residual force (\ref{model}) we have a significant improvement of the quality of fit compared to the zero residual hypothesis by $\delta\chi^2 >13$. This is demonstrated in Fig. \ref{chiosc} (right panel) where we show the minimized $\chi^2$ as a function of the spatial wavelength $\lambda$ of the macroscopic oscillating force (\ref{solmeoscilforce}). The depth of the best fit $\chi^2$ minimum is $\delta \chi^2 >13$ and is obtained for $\lambda\simeq 33\mu m$ which is almost the same value $\lambda\simeq 35\mu m$ of  the plain harmonic force parametrization (\ref{model})  shown on the left panel\footnote{The left panel of Fig. \ref{chiosc} is identical with Fig. \ref{chi2all} but we show it here again for easier comparison with the corresponding Figure obtained using the full source integral (\ref{solmeoscilforce}) rather than the simple parametrization (\ref{model}).}. 

In Fig. \ref{bfitnew} (right panel) we show the best fit macroscopic oscillating force parametrization (\ref{solmeoscilforce}) superposed with the SOLME residual force data. For comparison we also show the corresponding best fit of the plain harmonic parametrization (\ref{model}). The quality of fit (value of $\chi^2$) is almost identical despite the fact that the right panel shows the full source integral best fit parametrization where the oscillation amplitude decreases slowly with $z$.

\subsection{Oscillating Source Integral in Cartesian Coordinates}

\begin{figure}[b]
\centering
\vspace{0cm}\rotatebox{0}{\vspace{0cm}\hspace{0cm}\resizebox{0.49\textwidth}{!}{\includegraphics{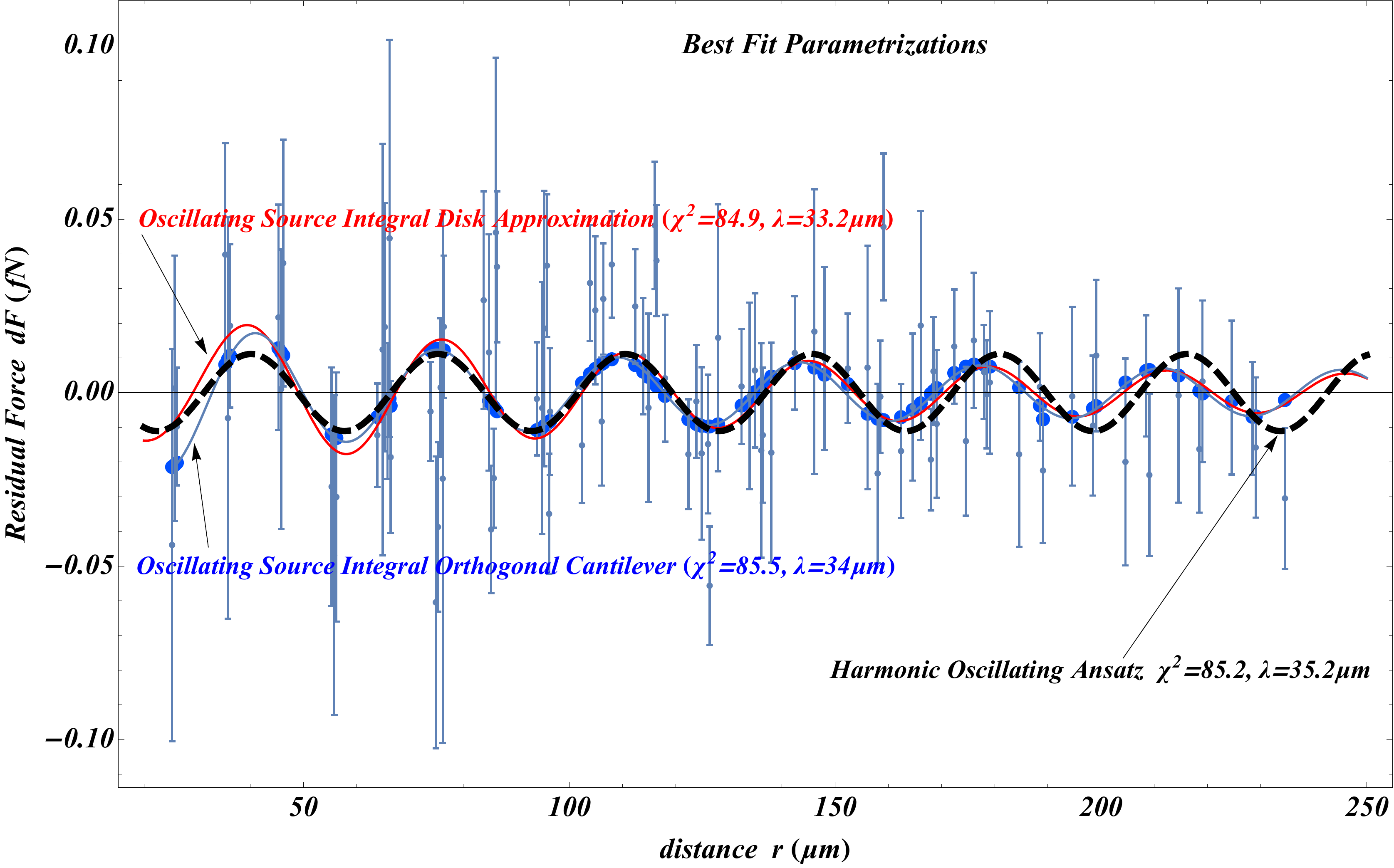}}}
\caption{The best fit source integrals for cylindrical (\ref{solmeoscilforce}) and othogonal (\ref{oscilforcecart}) cantilever along with the best fit plain harmonic force residual ansatz (\ref{model}).The three best fit parametrizations are very similar leading to practically the same quality of fit.} \label{altsourceints}
\end{figure}

In order to make the evaluation of the source integral analytically tractable we have approximated the orthogonal cantilever used in the SOLME by a cylindrical one of the same base area. The orthogonal cantilever used in the SOLME had dimensions $a\times b\times D =10 \mu m \times 500\mu m \times 2000\mu m$.   Had we kept the orthogonal geometry in the evaluation of the oscillating force source integral and rescaled with the dimension $a=10\mu m$ of the cantilever we would have to calculate the following source integral
\begin{widetext}
\be
F_{Oz}\left(\bar{z_0},\bar{\lambda},\theta\right)=Gm\rho a\times
\alpha_O \times \frac{\partial}{\partial \bar{z}_0}\int_{-1}^1
d\bar{x}\int_{-\bar{b}}^{\bar{b}} d\bar{y}
\int_{\bar{z}_0}^{\bar{z}_0+\bar{D}}\frac{\cos{\left(\frac{\sqrt{\bar{x}^2+\bar{y}^2+\bar{z}^2}}{\bar\lambda}+\theta\right)}}{\sqrt{\bar{x}^2+\bar{y}^2+\bar{z}^2}}
d\bar{z}\label{oscilforcecart}\ee
\end{widetext}
which in contrast to the cylindrical geometry is not analytically tractable. Using a numerical approach we have evaluated the source integral (\ref{oscilforcecart}) at the distances of the datapoints and confirmed that a similar quality of fit can be obtained using the orthogonal source integral (\ref{oscilforcecart}) as with the cylindrical analytic source integral (\ref{solmeoscilforce}) for the same spatial wavelength. Thus our result for the existence of the oscillating signal is robust and insensitive to the particular geometry used for the evaluation of the source integral. This is demonstrated in Fig. \ref{altsourceints} where we show the best fit source integrals for cylindrical (\ref{solmeoscilforce}) and othogonal (\ref{oscilforcecart}) cantilever along with the best fit plain harmonic force residual ansatz (\ref{model}). Clearly the three best fit parametrizations are very similar leading to practically the same quality of fit ($\chi^2\simeq 85$) compared to the much lower quality of fit for the zero residual hypothesis and the Yukawa or power law residuals ($\chi^2 \simeq 98$).

\section{Conclusions-Discussion}
\label{sec:Section 4}

We have analysed and fit the Stanford Optically Levitated Microsphere Experiment (SOLME)  \cite{Rider:2016xaq} force residual data using a wide range of parametrizations including plain phenomenological ones and parametrizations obtained by evaluating source integrals based on simple functional forms. We have shown that monotonic parametrizations (Yukawa and power laws) are unable to improve the quality of fit of the null hypothesis (zero force residuals) at any significant level despite the introduction of a number of parameters ($\delta \chi^2 <1$). In contrast, oscillating parametrizations at the plain phenomenological level and at the level of source integral can impvove significantly the quality of fit compared to the null hypothesis ($\delta \chi^2 >13$). The statistical significance of this oscillating signal is at about $2\sigma$ level. 

The most probable cause of this signal is a systematic effect caused by the non-Gaussian tails of the laser beam whose pressure levitates the microsphere.  Due to diffraction, the intensity of these non-Gaussian tails has a periodic oscillation, which can mimic a spatially oscillating force signal. Thus the detected signal can only be used as an upper bound to physically interesting new forces of sub-mm oscillating nature. The amplitude $\alpha$ of such an oscillating force background with spatial wavelength $\lambda\simeq 35\mu m$ is bounded at the $2\sigma$ level as $\alpha<0.3 \times 10^{-17}N$. This bound is phenomenological and applicable to the conditions and geometry of the SOLME.

If the origin of this signal is assumed to be gravitational through a modified Newtonian potential of the form of eq. (\ref{vosc}) the bounds obtained on the parameter $\alpha_O$ are particularly weak ($\alpha_O < 10^7$) due to the partially shielded electrostatic backgrounds that limit the sensitivity of the experiment in measuring gravitational forces between the cantilever and the microsphere.

More interesting bounds on fundamental fifth force parameters could be obtained if the origin of the signal is assumed to be non-gravitational. In particular, it is plausible that a chameleon potential with multiple extrema can lead to a spatially oscillating fifth force which is screened in regions of high density via the chameleon mechanism. Consider for a example \cite{Khoury:2003rn}  the chameleon field profile around a spherical object of radius $R_c$ and density $\rho(r)$. The profile of the chameleon field which acts also as a potential for the chameleon fifth force is determined by the equation

\begin{equation}
\frac{d^2\phi}{dr^2} + \frac{2}{r}\frac{d\phi}{dr} = V_{,\phi} +
\frac{\beta}{M_{Pl}}\rho (r) e^{\beta\phi/M_{Pl}}\,,
\label{camprof}
\end{equation}
where $\beta$ is a parameter, $M_{Pl}$ is the Planck mass scale and $V(\phi)$ is the chameleon field self-interaction potential. The density profile may be approximated as
\begin{equation}
\rho (r) = \left\{
\begin{matrix}
\rho_c\qquad {\rm for}\;\;\; r<R_c \cr
\rho_\infty\qquad {\rm for}\;\;\; r>R_c
\end{matrix}
\right. \,.
\label{densasym}
\end{equation}
Let $\phi_c$ and $\phi_\infty$ be the chameleon field value that minimizes the effective potential $V_{eff}$ defined as 
$V_{eff}(\phi) \equiv V(\phi) + \rho(r) e^{\beta_i\phi/M_{Pl}}$
for $r<R_c$ and $r>R_c$, respectively.
For these field values we have \cite{Khoury:2003rn}
\begin{eqnarray}
\nonumber
& & V_{,\phi}(\phi_c) + \frac{\beta}{M_{Pl}}\rho_c e^{\beta\phi_c/M_{Pl}} = 0\; \\
& & V_{,\phi}(\phi_\infty) + \frac{\beta}{M_{Pl}}\rho_\infty e^{\beta\phi_\infty/M_{Pl}} = 0\,.
\end{eqnarray}
The screened chameleon fifth force is obtained from the profile solution of eq. (\ref{camprof}) with boundary conditions 
\begin{eqnarray}
\nonumber
& & \frac{d\phi}{dr} = 0 \qquad {\rm at}\;\;\;r=0\, \\
& & \phi\rightarrow \phi_\infty \qquad {\rm as}\;\;\; r\rightarrow \infty\,.
\label{boundcond}
\end{eqnarray}
It can be shown from the chameleon field action that the chameleon fifth force on a test particle of mass $M$ is of the form
\begin{equation}
\vec{F}_\phi=-\frac{\beta}{M_{Pl}} M\vec{\nabla}\phi\,.
\label{chamforce}
\end{equation}
Thus $\phi$ plays the role a potential for the chameleon induced fifth force. 

If the chameleon self interaction potential is monotonic between the central value $\phi_c$ and the asymptotic field value $\phi_\infty$ then $\phi(r)$ varies monotonically between its value $\phi_c$ in the center of the massive object and its asymptotic value $\phi_\infty$ which is approached exponentially fast in the exterior of the massive body. We thus obtain the usual screened fifth force obtained from the gradient of $\phi(r)$ which is maximized around a thin shell at the borderline of the massive object and goes rapidly to 0 in the intrerior and in the exterior of the object with significantly larger mass in the interior (screened region).

If on the other hand there are multiple extrema of the potential $V(\phi)$ in the range between the central value $\phi_c$ and the asymptotic field value $\phi_\infty$, then eq. (\ref{camprof}) implies that these extrema may be inherited to the chameleon field profile around the massive object. Thus from eq. (\ref{chamforce}) these multiple extrema may induce localized sub-mm spatial oscillations of the chameleon induced screened fifth force. A similar behavior may be obtained if the exponential conformal coupling to the density $e^{\beta\phi/M_{Pl}}$ is replaced by an oscillating function. The detailed investigation of this class of fifth forces and their signature in the SOLME data is an interesting extension of the present project.

{\bf Supplemental Material:} The numerical analysis Mathematica files used for the construction of the figures and the derivations of the source integrals may be found in \href{http://leandros.physics.uoi.gr/solme/}{this url.}

\section*{Acknowledgements}
We thank the authors of Ref. \cite{Rider:2016xaq} and especially Prof. David Moore and Prof. Giorgio Gratta for providing their dataset, for confirming the presence of the detected signal in their data and for useful comments regarding the possible origin of this oscillating signal.

\section*{Appendix}
In Table \ref{tab} we show the dataset used in our maximum likelihood analysis. The dataset includes the distance between the center of the microsphera and the origin of a cartesian coordinate system which located in the center of the front side of the cantilever (see Fig. $1$ of \cite{Rider:2016xaq}), the residual force (the difference between the measured force $F$ and the electrostatic background $F_B$), the corresponding $1\sigma$ error and the number of the experiment (microsphera). The dataset was kindly provided by the authors of Ref. \cite{Rider:2016xaq}) after our request.

\begin{longtable}{ | c | c | c | c |}
\caption{The residual force 96 datapoints used for the $\chi^2$ analysis.}\label{tab}\\
\hline
    $r \; (\mu m)$ &  $F-F_B$ ($fN$)& $1\sigma \; (F-F_B)$   & Microsphera \\ \hline
 26.2 & -0.0098 & 0.017 & I \\
 36.2 & 0.0193 & 0.0235 & I\\
 46.2 & 0.0373 & 0.0355 & I\\
 56.2 & -0.0301 & 0.0359 & I\\
 66.2 & 0.0445 & 0.0572 & I\\
 66.4 & -0.0186 & 0.0219 & I\\
 76.2 & -0.0248 & 0.0762 & I\\
 76.4 & 0.019 & 0.0205 & I\\
 86.2 & 0.0461 & 0.0504 & I\\
 86.4 & 0.0363 & 0.0218 & I\\
 96.2 & -0.035 & 0.0173 & I\\
 96.4 & -0.0055 & 0.0183 & I \\
 102.4 & -0.0152 & 0.0166 & I \\
 106.4 & 0.027 & 0.016 & I\\
 112.4 & 0.0248 & 0.0165 & I\\
 116.4 & 0.038 & 0.016 & I\\
 122.4 & -0.0178 & 0.0158 & I\\
 126.4 & -0.0557 & 0.0171 & I\\
 132.4 & 0.0017 & 0.0165 & I\\
 136.4 & -0.0123 & 0.0195 & I\\
 142.4 & 0.0114 & 0.0164 & I\\
 152.4 & 0.007 & 0.0158 & I\\
 158.5 & -0.0012 & 0.0162 & I\\
 162.4 & -0.0169 & 0.0193 & I\\
 168.5 & 0.0061 & 0.0156 & I\\
 172.4 & 0.0133 & 0.0165 & I\\
 178.5 & -0.0005 & 0.0156 & I\\
 188.5 & 0.0016 & 0.0155 & I\\
 198.5 & -0.0095 & 0.0202 & I\\
 208.5 & 0.0048 & 0.0175 & I\\
 218.5 & -0.0163 & 0.0183 & I\\
 228.5 & -0.0075 & 0.0162 & I\\
 25.3 & -0.0439 & 0.0565 & II\\
 35.3 & 0.0398 & 0.0321 & II\\
 45.3 & 0.0217 & 0.0325 & II\\
 55.3 & -0.0271 & 0.0344 & II\\
 64.9 & 0.0124 & 0.0593 & II\\
 65.3 & 0.0189 & 0.0358 & II\\
 74.9 & -0.0605 & 0.042 & II\\
 75.3 & -0.0388 & 0.0244 & II\\
 84.9 & 0.0116 & 0.0341 & II\\
 85.3 & -0.0395 & 0.0184 & II\\
 94.9 & -0.0045 & 0.0364 & II\\
 95.3 & 0.0185 & 0.0397 & II\\
 104.9 & 0.0237 & 0.0213 & II\\
 106.1 & -0.0083 & 0.0185 & II\\
 114.9 & -0.0044 & 0.0271 & II\\
 116.1 & 0.0482 & 0.0183 & II\\
 124.9 & -0.0175 & 0.0249 & II\\
 126.1 & -0.0149 & 0.02 & II\\
 134.9 & 0.0064 & 0.0222 & II\\
 136.1 & -0.0167 & 0.0309 & II\\
 146.1 & 0.0176 & 0.041 & II\\
 156.1 & 0.0072 & 0.0351 & II\\
 159.1 & 0.0478 & 0.0211 & II\\
 166.1 & 0.0193 & 0.033 & II\\
 169.1 & -0.009 & 0.0213 &II\\
 176.1 & 0.015 & 0.0195 & II\\
 179.1 & 0.0029 & 0.0206 & II\\
 189.1 & -0.0225 & 0.0209 & II \\
 199.1 & 0.0107 & 0.0218 & II\\
 209.1 & -0.0238 & 0.0233 & II\\
 219.1 & 0.0032 & 0.0233 & II\\
 229.1 & -0.0158 & 0.0202 & II\\
 25.8 & 0.0012 & 0.0383 & III\\
 35.8 & -0.0074 & 0.0579 & III\\
 45.8 & 0.001 & 0.0402 & III\\
 55.8 & -0.0469 & 0.0461 & III\\
 63.9 & -0.0123 & 0.0149 & III\\
 65.8 & -0.0053 & 0.0196 & III\\
 73.9 & -0.0055 & 0.0143 & III\\
 75.8 & 0.0015 & 0.02 & III\\
 83.9 & 0.0266 & 0.0314 & III\\
 85.8 & -0.0247 & 0.0143 & III\\
 93.9 & -0.0018 & 0.0163 & III\\
 95.8 & 0.0366 & 0.0206 & III\\
 103.9 & 0.0316 & 0.0167 & III\\
 108. & 0.0369 & 0.0153 & III\\
 113.9 & 0.0105 & 0.0168 & III\\
 118. & 0.0041 & 0.0183 & III\\
 123.9 & -0.0025 & 0.0163 & III\\
 128. & 0.0158 & 0.0385 & III\\
 133.9 & -0.001 & 0.0269 & III\\
 138. & -0.0173 & 0.0317 & III\\
 148. & 0.0098 & 0.0264 & III\\
 158. & -0.0233 & 0.0258 & III\\
 164.6 & -0.0042 & 0.0212 & III\\
 168. & -0.0193 & 0.0146 & III\\
 174.6 & -0.014 & 0.0215 & III\\
 178. & 0.0059 & 0.0136 & III\\
 184.6 & -0.0178 & 0.0267 & III\\
 194.6 & -0.0011 & 0.0257 & III\\
 204.6 & -0.02 & 0.0298 & III\\
 214.6 & -0.0008 & 0.0309 & III\\
 224.6 & -0.0015 & 0.0222 & III\\
 234.6 & -0.0305 & 0.0204 & III\\
\hline
\end{longtable}

\raggedleft
\bibliography{oscillations}

\end{document}